\documentclass[11pt]{elsarticle}
\usepackage{amsmath,amssymb,amsthm}
\usepackage{mathrsfs}
\usepackage[bookmarks=true,
  bookmarksnumbered=true,
  bookmarksopen=true,
  colorlinks,
  pdfborder=001,
linkcolor=black]{hyperref}
\usepackage{bm}
\usepackage{array}
\usepackage{float}
\usepackage{color}
\usepackage{lineno}
\usepackage{subcaption}
\usepackage{stmaryrd}
\usepackage{extarrows}
\usepackage{subcaption}
\usepackage{multirow}
\usepackage{booktabs}
\newtheorem{rmk}{Remark}[section]
\newproof{pf}{Proof}
\numberwithin{equation}{section}
\numberwithin{figure}{section}
\numberwithin{table}{section}
\allowdisplaybreaks

\newcommand\dd{\mathrm{d}}

\newcommand\bF{\bm{F}}
\newcommand\bv{\bm{v}}
\newcommand\bU{\bm{U}}
\newcommand\bW{\bm{W}}

\newcommand\vx{v_1}
\newcommand\vy{v_2}
\newcommand\vz{v_3}
\newcommand\Bx{B_1}
\newcommand\By{B_2}
\newcommand\Bz{B_3}
\newcommand\pt{p_\text{tot}}
\newcommand\BB{\abs{\bm{B}}^2}

\newcommand\vB{\bv\cdot\bm{B}}

\newcommand\pd[2]{\dfrac{\partial {#1}}{\partial {#2}}}
\newcommand\abs[1]{\lvert #1 \rvert}

\begin{document}
%\linenumbers

\begin{frontmatter}

  \title{An analytical solution of the isentropic vortex  problem  in the  special relativistic magnetohydrodynamics}

  \author{Junming Duan}
  \ead{duanjm@pku.edu.cn}
  \author{Huazhong Tang\corref{cor1}}
  \ead{hztang@math.pku.edu.cn}
  \address{Center for Applied Physics and Technology, HEDPS and LMAM,
  School of Mathematical Sciences, Peking University, Beijing 100871, P.R. China}
  \cortext[cor1]{Corresponding author. Fax:~+86-10-62751801.}

  \begin{abstract}
  The isentropic vortex  problem is frequently solved to test the accuracy of
  numerical methods and verify corresponding  code. Unfortunately, its existing solution was derived in the relativistic magnetohydrodynamics by numerically solving an ordinary differential equation.
    This note provides an analytical solution of the 2D isentropic vortex problem
     with   explicit algebraic expressions in the  special relativistic hydrodynamics and magnetohydrodynamics and extends it to the 3D case.
%    the first analytical isentropic vortex test problem \emph{with algebraic expressions} for the 2D and 3D special relativistic magnetohydrodynamic (RMHD) equations,
%    without any integral or solving equations, which is useful for code verification.
   % with algebraic expressions
  \end{abstract}


\end{frontmatter}

\section{Introduction}
The relativistic description for the fluid dynamics
 at nearly the speed of light should be considered
 in investigating  the astrophysical
phenomena  from stellar to galactic scales, e.g., coalescing neutron stars, core
collapse supernovae,  active galactic nuclei, superluminal jets, the formation of black holes,
and gamma-ray bursts etc.
 In the rest laboratory frame, the 2D and 3D special relativistic magnetohydrodynamic (RMHD) equations
 can be cast into
\begin{equation}\label{eq:RMHDdiv1}
	\pd{\bU}{t}+\sum_{k=1}^d \pd{\bF_k(\bU)}{x_k}=0,~d=2,3,
\end{equation}
with the divergence-free constraint on the magnetic field
\begin{equation}\label{eq:divB}
	\sum_{k=1}^d \pd{B_k}{x_k}=0,
\end{equation}
where $\bU$ and $\bF_k$ are respectively
the conservative variable vector and the flux vector in the $x_k$-direction and defined by
\begin{equation}\label{eq:RMHDdiv2}
	\begin{aligned}
		&\bU=(D,\bm{m},E,\bm{B})^\mathrm{T},\\
		&\bF_k=(Dv_k,\bm{m}v_k-B_k(\bm{B}/W^2+(\vB)\bv)+\pt\bm{e}_k,m_k,v_k\bm{B}-B_k\bv)^\mathrm{T}.
	\end{aligned}
\end{equation}
Here  $D=\rho W$,
$\bm{m}=(\rho hW^2+\BB)\bv-(\bv\cdot\bm{B})\bm{B}$ and $E=DhW-\pt+\BB$ are the mass,
momentum and energy densities, respectively,
 $\rho$, $\bv=(v_1,\cdots,v_d)$ and $\bm{B}=(B_1,\cdots,B_d)$ denote
  the rest-mass density, the velocities and the magnetic fields, respectively,
  $\bm{e}_k$ denotes the $k$-th row of the
$d\times d$ unit matrix,
 $W=1/\sqrt{1-\abs{\bv}^2}$ is the Lorentz factor,
$\pt$ denotes the total pressure containing the gas pressure $p$ and
the magnetic pressure $p_m=\frac12\left(\abs{\bm{B}}^2/W^2 + (\vB)^2\right)$,
and $h$ is the specific enthalpy defined by $h=1 + e + p/\rho$ with  the specific internal energy $e$.
The governing equations \eqref{eq:RMHDdiv1}-\eqref{eq:RMHDdiv2} need to be closed by the equation of state,
which  is restricted in this note to the perfect gas
\begin{equation}\label{eq:EOS}
p=(\Gamma-1)\rho e,
\end{equation}
with the adiabatic index $\Gamma\in(1,2]$.
Setting $\bm{B} = \bm{0}$ in the RMHD equations leads to the corresponding relativistic hydrodynamic (RHD) equations.

The  system \eqref{eq:RMHDdiv1}-\eqref{eq:RMHDdiv2} becomes much more complicated than the Euler equations in gas dynamics due to the relativistic effect, so its analytic treatment is very challenging.
Numerical simulation is a powerful way to help us  better understand the physical mechanisms in the RHDs and RMHDs.
As a first step, the accuracy test should be conducted to verify the  convergence rate of the numerical schemes.
Usually, the smooth  test of a sine wave propagation with a constant density or pressure is  considered,
and its multidimensional version is implemented by conducting such test in an oblique direction on a multidimensional mesh.
However, due to possible pseudo cancellation of the leading error terms, sometimes such test may cover up the true error \cite{Balsara2004Second}, thus it is desirable to design some genuinely 2D and 3D test problems.
In \cite{Balsara2000Monotonicity,Balsara2004Second}, the smooth vortex problems with algebraic expressions are constructed for the compressible Euler equations and magnetohydrodynamic equations, respectively.
For the RHD and RMHD equations, an isentropic vortex problem is constructed in \cite{Balsara2016A},
where an ordinary differential equation (ODE) should be integrated numerically to obtain the initial solutions at each given grid point, which is not convenient.
In \cite{Ling2019Physical}, the analytic solution of the isentropic vortex problem with the algebraic expression is given for the RHD equations and has been used to test the accuracy of
the high-order accurate entropy conservative and stable schemes in \cite{Duan2020RHD,Duan2021RHDMM}.
This note aims at deriving  an analytical solution of the isentropic vortex problem \emph{with explicit algebraic expressions} for the 2D and 3D RMHD equations \eqref{eq:RMHDdiv1}-\eqref{eq:RMHDdiv2}.

%The note is organized as follows.
%Section \ref{section:Construction} introduces the entropy conditions for the RHD equations in Cartesian and curvilinear coordinates.
%Several 2D and 3D numerical experiments are conducted in Section \ref{section:Num} to validate the efficiency and the ability of our schemes in capturing the sharp transitions or discontinuities.
%Section \ref{section:Conclusion}  concludes the work with final remarks.
%

\section{Review of the existing isentropic vortex problem}\label{section:Review}

This section reviews the 2D isentropic vortex problem proposed in \cite{Balsara2016A}.
The computational domain is $[-R,R]\times[-R,R]$ with the periodic boundary conditions,
and a 2D \emph{steady} isentropic vortex is first constructed in its own rest frame $S$ with the coordinates $(t,\bm{x})$ and $\bm{x}=(x_1,x_2,x_3)$.

In cylindrical coordinates $(r,\theta,z)$ with $r=\sqrt{x_1^2+x_2^2}$, $\theta=\arctan(x_2/x_1)$,
and $z=x_3$, the RMHD equations \eqref{eq:RMHDdiv1}-\eqref{eq:RMHDdiv2}
becomes
%using
%\begin{align*}
%	\dfrac{\partial}{\partial x_1} &= \dfrac{\partial}{\partial r}\pd{r}{x_1} + \dfrac{\partial}{\partial \theta}\pd{\theta}{x_1} = \cos\theta\dfrac{\partial}{\partial r} - \dfrac{\sin\theta}{r}\dfrac{\partial}{\partial \theta}, \\
%	\dfrac{\partial}{\partial x_2} &= \dfrac{\partial}{\partial r}\pd{r}{x_2} + \dfrac{\partial}{\partial \theta}\pd{\theta}{x_2} = \sin\theta\dfrac{\partial}{\partial r} + \dfrac{\cos\theta}{r}\dfrac{\partial}{\partial \theta}
%\end{align*}
%and
%\begin{equation*}
%	a_1 = \cos\theta a_r - \sin\theta a_\theta,~
%	a_2 = \sin\theta a_r + \cos\theta a_\theta,~
%	a_3 = a_z,
%\end{equation*}
%key!!!
%            \pd{\bm{m}}{t} + \sum_{k=1}^3 \pd{\bm{v}B_k}{x_k} = 0
%==>     \pd{m_r}{t} + \nabla \pd{v_r\bm{B}} = \dfrac{v_\theta B_\theta}{r}
%==>     \pd{m_\theta}{t} + \nabla \pd{v_\theta\bm{B}} = - \dfrac{B_\theta v_r}{r}
%absorb RHS:   \pd{m_\theta}{t} + \nabla^r \pd{v_\theta\bm{B}} = - \dfrac{B_\theta v_r - B_r v_\theta}{r}, the flux tensor is symmetric, so RHS vanishes in the end
%one has
\begin{equation}
\begin{aligned}
	&\pd{D}{t}+\nabla\cdot(D\bv)=0,  \\
	&\pd{m_r}{t}+\nabla\cdot\left[m_r\bv
	-\left(B_r/W^2+(\vB)v_r\right)\bm{B} \right]+\pd{\pt}{r}=\dfrac{m_\theta v_\theta}{r}
	-\left({B_\theta}/{W^2} + (\vB)v_\theta\right)\dfrac{B_\theta}{r}, \\
	&\pd{m_\theta}{t}+\nabla^r\cdot\left[m_\theta\bv
	-\left(B_\theta/W^2+(\vB)v_\theta\right)\bm{B} \right] + \dfrac{1}{r}\pd{\pt}{\theta}=0, \\
	&\pd{m_z}{t}+\nabla\cdot\left[m_z\bv
	-\left(B_z/W^2+(\vB)v_z\right)\bm{B} \right] + \pd{\pt}{z}=0, \\
	&\pd{E}{t} + \nabla\cdot \bm{m}=0, \\
	&\pd{B_r}{t} + \dfrac{1}{r}\pd{E_z}{\theta} - \pd{E_\theta}{z} = 0, \\
	&\pd{B_\theta}{t} + \pd{E_r}{z} - \pd{E_z}{r} = 0, \\
	&\pd{B_z}{t} + \dfrac{1}{r}\pd{(rE_\theta)}{r} - \dfrac{1}{r}\pd{E_r}{\theta} = 0,
\end{aligned}\label{eq:RMHD_cyl}\end{equation}
where $\bm{E}=(E_r, E_\theta, E_z)=-\bv\times \bm{B}$ is the electric vectors,
and the symbols $\nabla$ and $\nabla^r$ are respectively defined by
\begin{equation*}
	\nabla\cdot\bm{a}=\dfrac{1}{r}\pd{(ra_r)}{r}+\dfrac{1}{r}\pd{a_\theta}{\theta} +\pd{a_z}{z},\quad
	\nabla^r\cdot\bm{a}=\dfrac{1}{r^2}\pd{(r^2a_r)}{r}+\dfrac{1}{r}\pd{a_\theta}{\theta} + \pd{a_z}{z},
\end{equation*}
here $a_r,a_\theta,a_z$ are the radial, angular and $z$-component of a vector $\bm{a}$, respectively.
If assuming that the rest-mass density and the pressure are univariate functions of the radius $r$
\begin{equation}\label{eq:RhoPre}
	\rho=\rho(r),~p=\rho^\Gamma,~
\end{equation}
and making an ansatz for the velocities and the magnetic fields as follows
\begin{equation}\label{eq:VelMag}
	(\vx, \vy)=(-x_2, x_1)f(r),~
	(\Bx, \By)=(-x_2, x_1)g(r),~
	\vz=\Bz=0,
\end{equation}
then
\begin{equation}\label{eq:RTComp}
	v_r = 0,~v_\theta=rf,~B_r=0,~B_\theta=rg,
\end{equation}
where $f(r)$ and $g(r)$ are univariate functions of $r$, determined later.
Now it is easy to see that the divergence constraint \eqref{eq:divB} and the equations in \eqref{eq:RMHD_cyl} hold automatically except for the 2nd, which becomes
\begin{equation}\label{eq:PressureODE}
	r\dfrac{\dd \left(p + \frac12r^2g^2 \right)}{\dd r}=\rho hW^2r^2f^2 - r^2g^2,
\end{equation}
where $p$, $f$, $g$ are unknown.

In \cite{Balsara2016A}, the authors chose
$f(r)=v_{\text{max}}^{\theta}\exp((1-r^2)/2)$ and $g(r)=B_{\text{max}}^{\theta}\exp((1-r^2)/2)$,
with the parameters $v_{\text{max}}^{\theta}=0.7,~B_{\text{max}}^{\theta}=0$ or $v_{\text{max}}^{\theta}=B_{\text{max}}^{\theta}=0.7$ for the RHD or RMHD cases, respectively,
thus the velocities and magnetic fields diminish rapidly when $r$ increases, which will not cause boundary effects when the periodic boundary conditions are used.
Then the ODE \eqref{eq:PressureODE} with the initial condition $p|_{r=0}=1$ is solved numerically to obtain the pressure $p$ and then the rest-mass density $\rho$ by the isentropic condition.
In other words, the radial ODE \eqref{eq:PressureODE} should be integrated numerically  from the center of the vortex ($r=0$) to set the initial conditions of the isentropic vortex problem at each given grid point.
Some further remarks about the solution of the ODE \eqref{eq:PressureODE} will be given in Remark \ref{rmk:ODE}.

\begin{figure}[!ht]
	\centering
	\begin{subfigure}[b]{0.48\textwidth}
		\centering
		\includegraphics[width=0.8\textwidth]{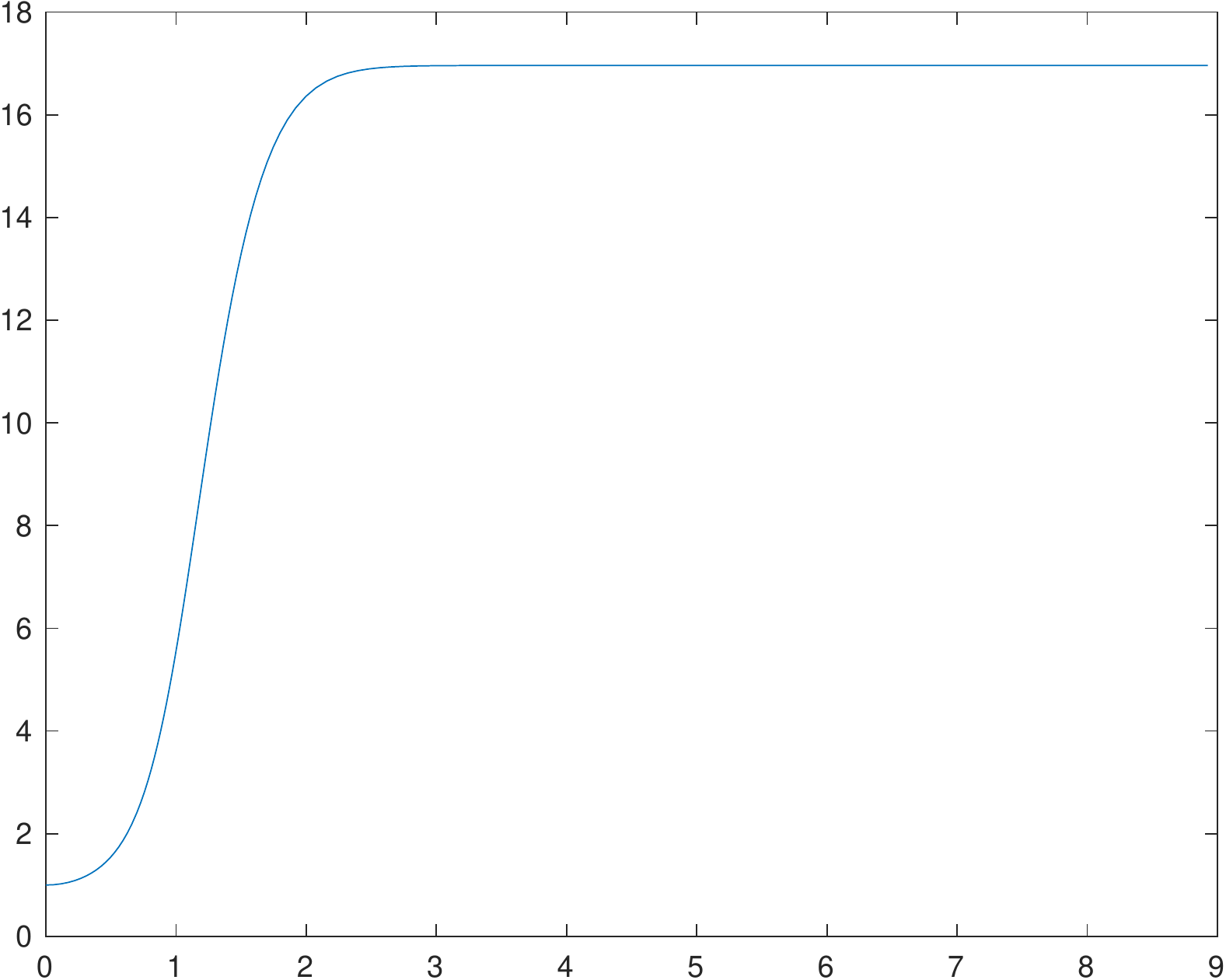}
		\caption{$p$}
	\end{subfigure}
	\begin{subfigure}[b]{0.48\textwidth}
		\centering
		\includegraphics[width=0.8\textwidth]{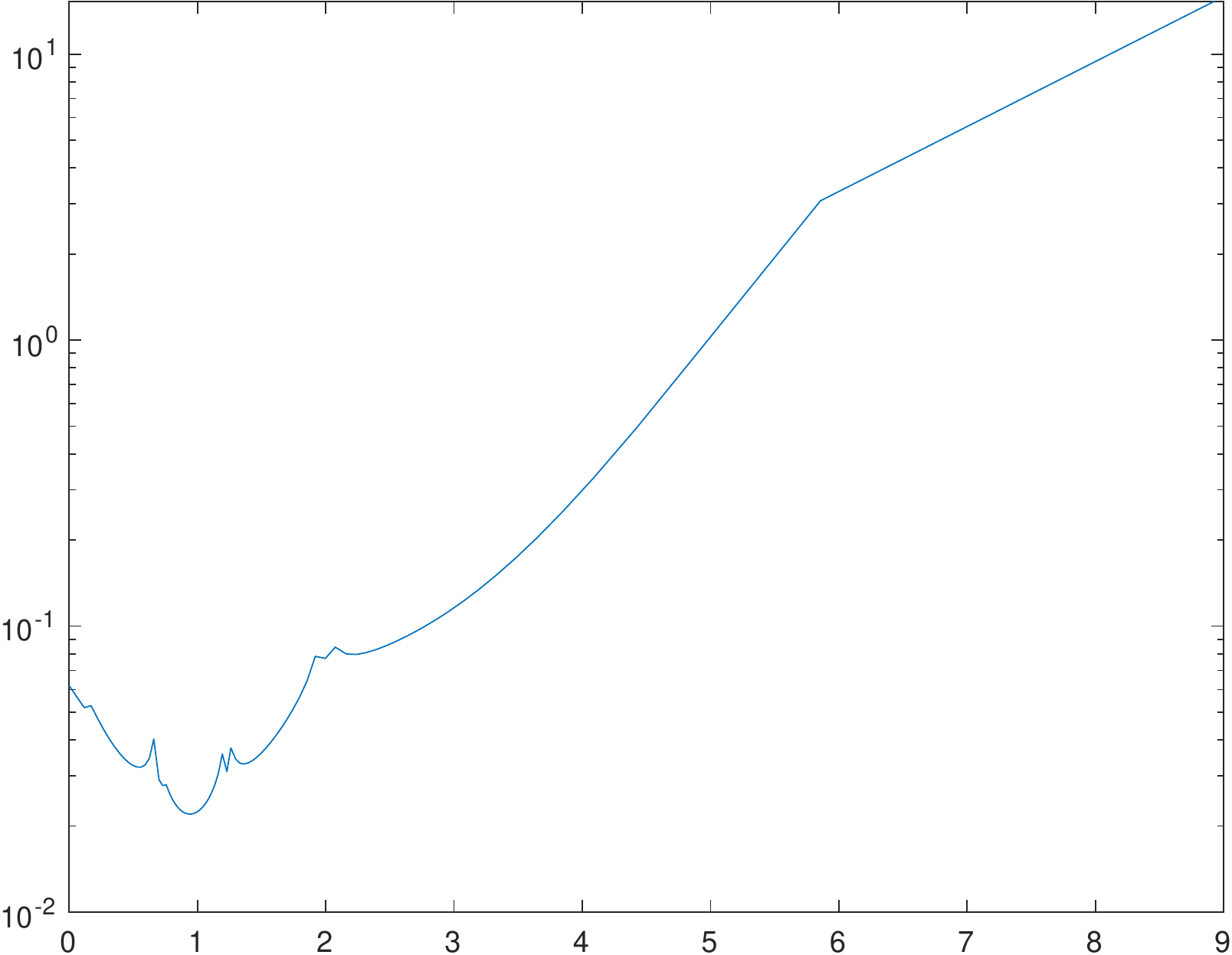}
		\caption{$\Delta r$}
	\end{subfigure}
	\caption{The isentropic vortex problem in \cite{Balsara2016A}: the pressure $p$ and the step size $\Delta r$ %with respect to the radius $r$
		obtained by solving the ODE \eqref{eq:PressureODE} with the variable order Runge-Kutta method \cite{Cash1990}.}
	\label{fig:ODE_r}
\end{figure}

\begin{rmk}\label{rmk:ODE}\rm
In \cite{Balsara2016A}, the authors suggested that ``the run of density and pressure for the vortices should be tabulated on a very fine one-dimensional radial mesh. Typically, this radial mesh should have resolution that is \emph{much finer} than the two-dimensional mesh on which the problem is computed."
We find that a more efficient way is to sort all the corresponding radius $r$, where the initial value is needed as a vector, and then integrate ODE \eqref{eq:PressureODE} to those sorted radius.
Since there is a steep transition of the pressure $p$ with respect to the radius $r$ near $r\in(1,2)$, see Figure \ref{fig:ODE_r}, %where the horizontal coordinate-axis is $r$,
a good ODE solver is the variable order Runge-Kutta method \cite{Cash1990},
which adaptively adjusts the step size by using the difference of the fourth-order and fifth-order solutions as an error estimator, thus the efficiency of solving the ODE \eqref{eq:PressureODE} can be improved.
From Figure \ref{fig:ODE_r}, one can see that  the step size varies from $\mathcal{O}(10^{-2})$ to $\mathcal{O}(1)$, so that using adaptive step size is efficient.
Nevertheless, it is still desirable to obtain an analytical solution
of the ODE \eqref{eq:PressureODE} with \emph{algebraic expressions}%\footnote{Basic arithmetic operations and rational exponent are involved.}
, which will be more convenient for the researchers.
But, it is not trivial.
\end{rmk}

\begin{rmk}\rm
	For the non-relativistic hydrodynamics, the ODE \eqref{eq:PressureODE} with the ansatz $(\vx, \vy)=(-x_2, x_1)v_{\text{max}}^{\theta}\exp((1-r^2)/2)$ and $p=\rho^\Gamma$ reduces to
	\begin{equation*}
		\dfrac{\dd p}{\dd r}=p^{1/\Gamma}(v_{\text{max}}^{\theta})^2r\exp(1-r^2).
	\end{equation*}
	One can solve it analytically, and obtain
	\begin{equation}\label{eq:EulerRho}
		p = \left(C_1 - \dfrac{(\Gamma-1)(v_{\text{max}}^{\theta})^2}{2\Gamma}\exp(1-r^2)\right)^{\Gamma/(\Gamma-1)},
	\end{equation}
	which is just the case given in \cite{Balsara2000Monotonicity} with {$C_1=1, v_{\text{max}}^{\theta}={5}/{2\pi}$. }
\end{rmk}

A time-dependent vortex moving with the velocity $\bm{w}=(w_1,w_2)^\mathrm{T}$ can be obtained by the Lorentz transformation.
Assume that a coordinate system $\widetilde{S}$ with the spacetime coordinates $(\widetilde{t},\widetilde{\bm{x}})$ and $\widetilde{\bm{x}}=(\widetilde{x}_1,\widetilde{x}_2,\widetilde{x}_3)$ is in motion relative to the coordinate system $S$ with a constant velocity $-\bm{w}$. The relation between the four-coordinates in $S$ and $\widetilde{S}$ is given by the Lorentz transformation as follows
\begin{equation*}
	\begin{pmatrix}
		t \\ x_1 \\ x_2 \\ x_3 \\
	\end{pmatrix}
	=\begin{pmatrix}
		\gamma & -\gamma w_1 & -\gamma w_2 & 0 \\
		-\gamma w_1 & 1+(\gamma-1) w_1^2/\abs{\bm{w}}^2 & (\gamma-1) w_1w_2/\abs{\bm{w}}^2 & 0 \\
		-\gamma w_2 & (\gamma-1) w_1w_2/\abs{\bm{w}}^2 & 1+(\gamma-1) w_2^2/\abs{\bm{w}}^2 & 0 \\
		0 & 0 & 0 & 1 \\
	\end{pmatrix}
	\begin{pmatrix}
		\widetilde{t} \\ \widetilde{x}_1 \\ \widetilde{x}_2 \\ \widetilde{x}_3 \\
	\end{pmatrix},
\end{equation*}
where $\gamma=1/\sqrt{1-\abs{\bm{w}}^2}$.
The scalar variables such as $\rho$ and $p$ are invariant
\begin{equation*}
	\rho(\widetilde{t}, \widetilde{\bm{x}}) = \rho(\widetilde{t}(t,\bm{x}), \widetilde{\bm{x}}(t,\bm{x})),~
	p(\widetilde{t}, \widetilde{\bm{x}}) = \rho(\widetilde{t}(t,\bm{x}), \widetilde{\bm{x}}(t,\bm{x})),
\end{equation*}
while the velocities in $\widetilde{S}$ are
\begin{align*}
	\widetilde{v}_1&=\dfrac{1}{\gamma\left( 1+w_1\vx+w_2\vy \right)}\left[\gamma w_1
	+ \left(1 + (\gamma-1)w_1^2/\abs{\bm{w}}^2\right)\vx
	+ (\gamma-1)w_1w_2/\abs{\bm{w}}^2\vy\right],\\
	\widetilde{v}_2&=\dfrac{1}{\gamma\left( 1+w_1\vx+w_2\vy \right)}\left[\gamma w_2
	+ (\gamma-1)w_1w_2/\abs{\bm{w}}^2\vx
	+ \left(1 + (\gamma-1)w_2^2/\abs{\bm{w}}^2\right)\vy \right],\\
	\widetilde{v}_3&=0.
\end{align*}
If denoting the electric field potential and the magnetic vector potential by $\Upsilon$ and $\bm{A}$, respectively,
then $\left(\Upsilon,\bm{A}\right)$ transforms like a four-coordinate.
Since $\bm{E}=-\bv\times\bm{B}=\bm{0}$ ($\bv\parallel\bm{B}$), one can simply set $\Upsilon=0$, and obtain $\widetilde{A}_3=A_3=B_0\exp((1-r^2)/2)$,~$\widetilde{A}_1=\widetilde{A}_2=0$.
Thus the magnetic fields in $\widetilde{S}$ are
\begin{align*}
	\widetilde{B}_1&=\pd{\widetilde{A}_3}{\widetilde{x}_2}=\pd{\widetilde{A}_3}{x_1}\pd{x_1}{\widetilde{x}_2}
	+\pd{\widetilde{A}_3}{x_2}\pd{x_2}{\widetilde{x}_2}
	= -(\gamma-1) w_1w_2/\abs{\bm{w}}^2 \By + \left(1 + (\gamma-1) w_2^2/\abs{\bm{w}}^2\right) \Bx,\\
	\widetilde{B}_2&=-\pd{\widetilde{A}_3}{\widetilde{x}_1}=-\pd{\widetilde{A}_3}{x_1}\pd{x_1}{\widetilde{x}_1}
	-\pd{\widetilde{A}_3}{x_2}\pd{x_2}{\widetilde{x}_1}
	=  \left(1 + (\gamma-1) w_1^2/\abs{\bm{w}}^2\right) \By -(\gamma-1) w_1w_2/\abs{\bm{w}}^2 \Bx,\\
	\widetilde{B}_3&=0.
\end{align*}
In practice, the computation is usually performed in the coordinate system ${S}$ by exchanging the roles of the two coordinate systems ${S}$ and $\widetilde{S}$.

\section{Analytical solution of the 2D isentropic vortex}\label{section:Construction}
%Similar to \cite{Ling2019Physical} and \cite{Balsara2016A}, a two-dimensional \emph{steady} isentropic vortex should be constructed in its own rest frame $S$.
To begin with, let us first consider the RHD case (i.e. $\bm{B}=\bm{0}$ or $g=0$),
which is helpful for us to deal with the RMHD case.
It is worth noting that the analytical solution of the 2D isentropic vortex for this case
has been provided in \cite{Ling2019Physical}  without the detailed derivation and used in \cite{Duan2020RHD,Duan2021RHDMM}.

 In the RHD case,  the ODE \eqref{eq:PressureODE} becomes
\begin{equation}\label{eq:PressureODE_RHD}
	r\dfrac{\dd p}{\dd r}=\rho hW^2r^2f^2,
\end{equation}
where $p$ and $f$ are unknown.
 The key point in deriving our analytical solution of the 2D isentropic vortex
 is to \emph{make ansatz for $\rho$ rather than $\bv$ or $f$}, such that the   ODE \eqref{eq:PressureODE_RHD} does not contain the derivative of the unknown variable $p$,
and then   reduces to an \emph{algebraic equation} of $f$, which can be solved analytically.
%{\color{red}
%	Specifically, make the following ansatz similar to the non-relativistic case
%	\begin{equation}\label{eq:RhoRHD}
%		\rho=(1-\sigma \exp(1-r^2))^{1/(\Gamma-1)},
%	\end{equation}
%	where $\sigma$ is a constant used to control the range of $\rho$.
%}
Specifically, make the following ansatz
\begin{equation}\label{eq:RhoRHD}
	\rho=(1-\sigma \exp(1-r^2))^\varsigma,
\end{equation}
where $\sigma$ is a constant used to control the range of $\rho$ such that $\rho>0$, and $\varsigma$ is a constant to be determined later.
Such ansatz  is suitable since it tends to unity rapidly when $r$ increases.
Moreover, it contains the density profile in  the non-relativistic case \eqref{eq:EulerRho} in \cite{Balsara2000Monotonicity}, where $\sigma={25(\Gamma-1)}/\left(8\Gamma\pi^2\right)$ and $\varsigma=1/(\Gamma-1)$.
%and $\exp(1-r^2)\approx 4.36\times 10^{-28}$ with $r=8$,
%thus \eqref{eq:RhoRHD} is an appropriate ansatz since it tends to unity rapidly when $r$ increases,

Substituting the first two equations in \eqref{eq:RTComp} and \eqref{eq:RhoRHD} into \eqref{eq:PressureODE_RHD} yields
\begin{equation*}
	2(\Gamma-1)\Gamma\sigma\varsigma \exp(1-r^2)=
	\left((\Gamma-1)(1-\sigma\exp(1-r^2))^{1-(\Gamma-1)\varsigma}+\Gamma (1-\sigma\exp(1-r^2))\right)\dfrac{f^2}{1-r^2f^2},
\end{equation*}
then one can get
\begin{equation*}
	f^2=\dfrac{2(\Gamma-1)\Gamma\sigma\varsigma \exp(1-r^2)}{2(\Gamma-1)\Gamma\sigma\varsigma \exp(1-r^2)r^2 + (\Gamma-1)(1-\sigma\exp(1-r^2))^{1-(\Gamma-1)\varsigma}+\Gamma (1-\sigma\exp(1-r^2))}.
\end{equation*}
If $\sigma\varsigma<0$, the numerator of $f^2$ is always negative, and the denominator tends to $2\Gamma-1>0$ when $r$ goes to infinity, hence $f^2<0$, which means $f$ is not well-defined.
Therefore, we have to consider the case of $\sigma\varsigma>0$.
Assume that  $\sigma>0$ and $\varsigma>0$,
to simplify the expression of $f^2$, we choose
\begin{equation}\label{eq:Index}
	\varsigma=1/(\Gamma-1),
\end{equation}
so that one has
\begin{equation}\label{eq:f_RHD}
	f=\sqrt{\dfrac{2\Gamma\sigma \exp(1-r^2)}{2\Gamma\sigma \exp(1-r^2)r^2 + 2\Gamma-1-\Gamma\sigma\exp(1-r^2)}}.
\end{equation}
%Moreover, it can be verified that $f < 1.08\times 10^{-11}$ when $r>5$ with $\sigma=0.2,\Gamma=5/3$, satisfying the restriction that the velocities vanish rapidly as $r$ increases.
It can be verified  from the above expression of $f$ that the velocities $v_1$ and $v_2$ vanish rapidly as $r$ increases.
To sum up, \eqref{eq:RhoPre}-\eqref{eq:VelMag} and \eqref{eq:RhoRHD}-\eqref{eq:f_RHD}
% with $\sigma=0.2,\Gamma=5/3$
gives algebraic expressions of the analytical solution of the 2D isentropic vortex problem  in the RHDs,
which  have been provided in \cite{Ling2019Physical}.
The case of $\sigma<0$ and $\varsigma<0$ may be similarly discussed by changing  $(\sigma,\varsigma)$ as
$(-\sigma,-\varsigma)$.

\begin{figure}[!ht]
	\centering
	\begin{subfigure}[b]{0.48\textwidth}
		\centering
		\includegraphics[width=.8\textwidth]{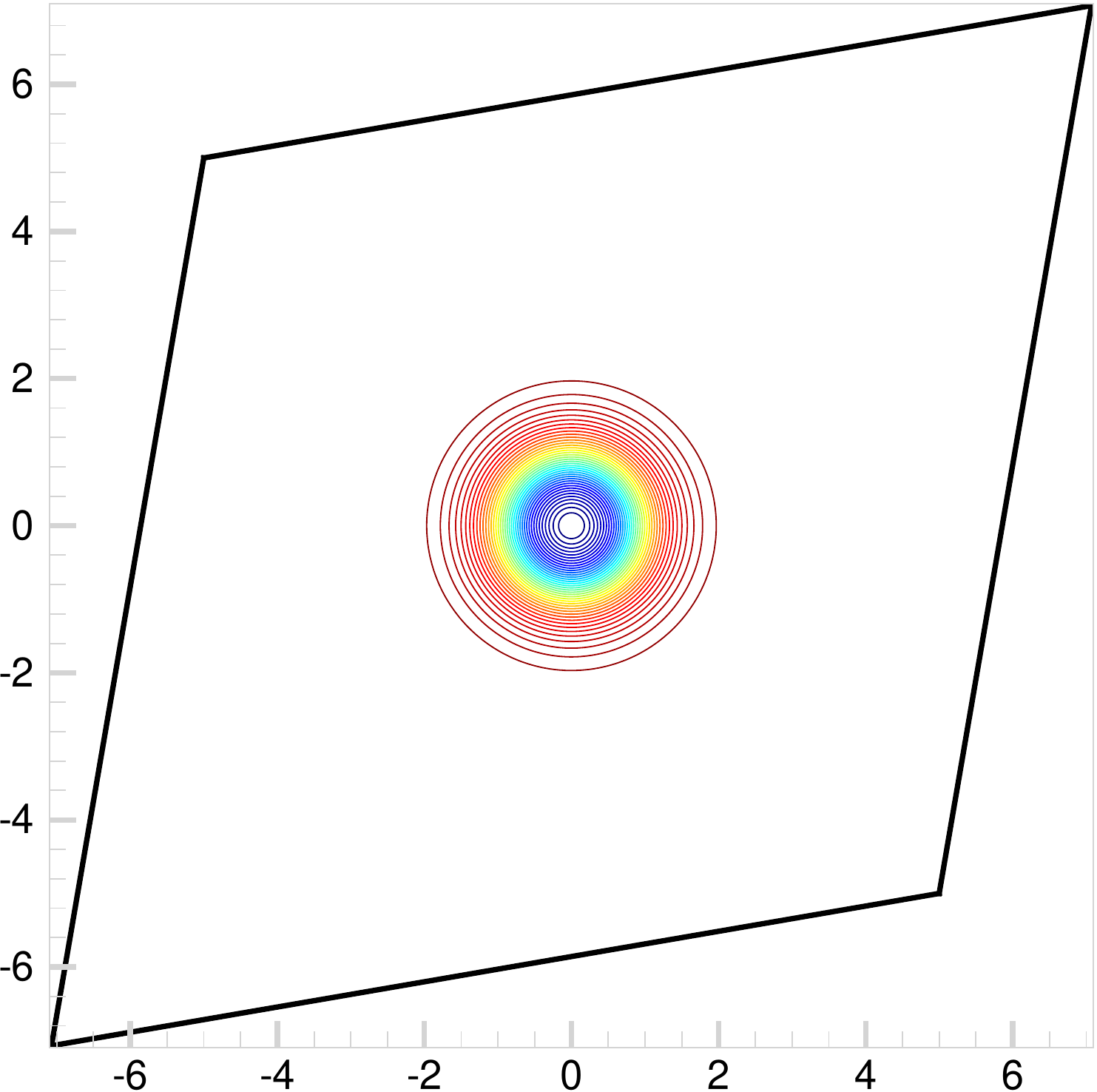}
	\end{subfigure}
	\begin{subfigure}[b]{0.48\textwidth}
		\centering
		\includegraphics[width=0.8\textwidth]{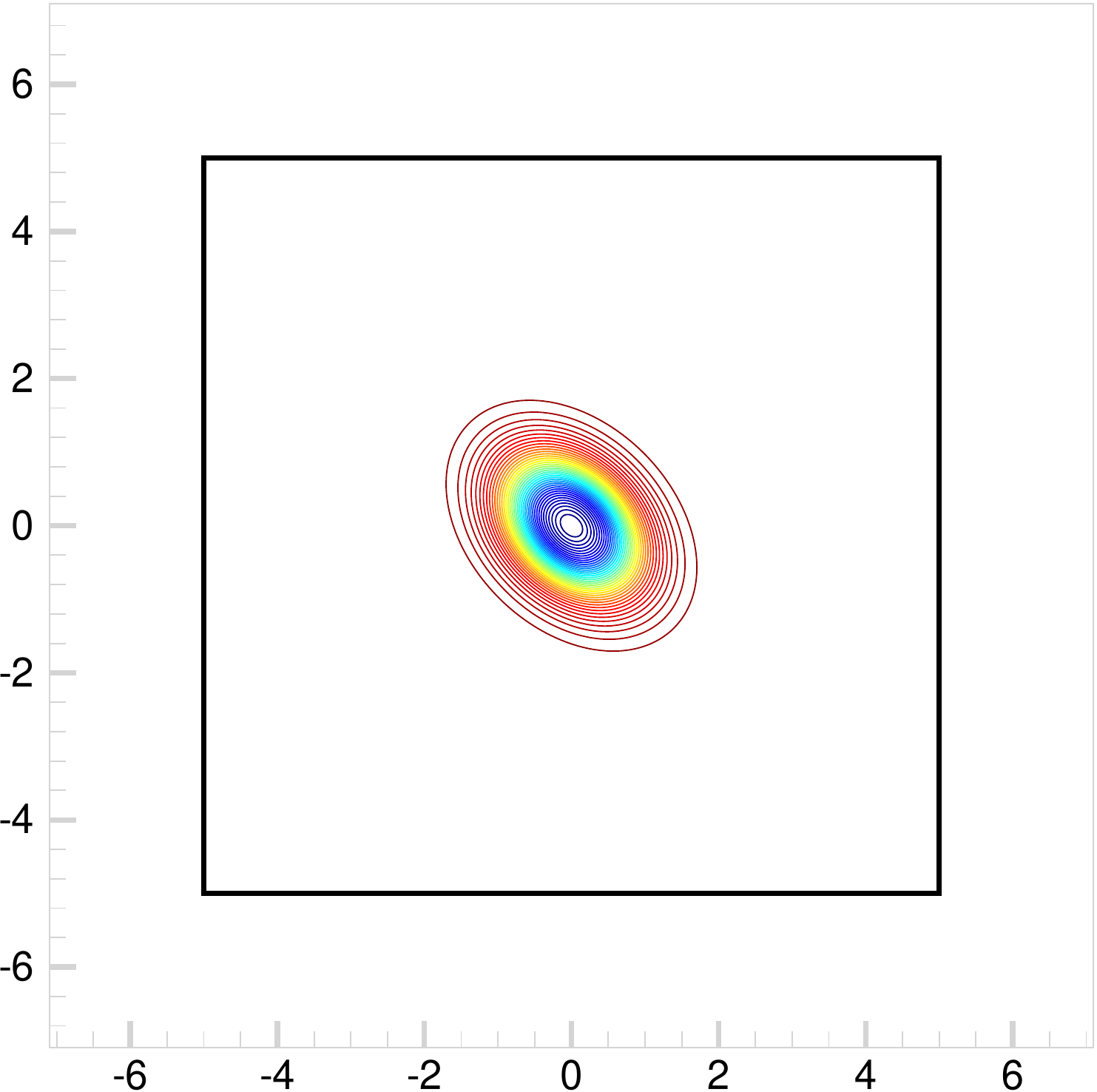}
	\end{subfigure}
	\caption{2D RMHD case: $40$ equally spaced contour lines of the initial $\rho$ in the rest frame and the moving frame with $R=5$ and $\bm{w}=(-0.5,-0.5)$.}
	\label{fig:domain}
\end{figure}

Now let's consider the RMHD case with non-zero magnetic fields.
Use the ansatz \eqref{eq:RhoRHD}-\eqref{eq:Index} for the rest-mass density and make an ansatz for the magnetic fields as follows
\begin{equation}\label{eq:MagRMHD}
  g=B_0\exp((1-r^2)/2).
\end{equation}
%
%Substituting \eqref{eq:RTComp} and \eqref{eq:RhoRHD}, and \eqref{eq:MagRMHD} into \eqref{eq:PressureODE} one can obtain
%\begin{align*}
%	2(\Gamma-1)\left[\Gamma\sigma\varsigma\exp(1-r^2)(1-\sigma\exp(1-r^2))^{\varsigma\Gamma-1}
%	+B_0^2\exp(2\mu(1-r^2)) - B_0^2r^2\mu\exp(2\mu(1-r^2))\right]= \\
%	\left((\Gamma-1)(1-\sigma\exp(1-r^2))^{\varsigma}+\Gamma (1-\sigma\exp(1-r^2)^{\varsigma\Gamma})\right)\dfrac{f^2}{1-r^2f^2},
%\end{align*}
%Taking $\mu=1/2,~\varsigma=1/\Gamma$, one has
%\begin{align*}
%	(\Gamma-1)\exp(1-r^2)\left[2\sigma+B_0^2(2-r^2)\right]= \\
%	\left((\Gamma-1)(1-\sigma\exp(1-r^2))^{1/\Gamma}+\Gamma (1-\sigma\exp(1-r^2))\right)\dfrac{f^2}{1-r^2f^2},
%\end{align*}
%thus
%\begin{equation}\label{eq:f2}
%	f^2=\dfrac{\kappa (\Gamma-1)\exp(1-r^2)}{\kappa r^2(\Gamma-1)\exp(1-r^2) + (\Gamma-1)\rho + \Gamma p},
%\end{equation}
%where
%\begin{equation}\label{eq:kappa}
%	\kappa = 2\sigma+B_0^2(2-r^2).
%\end{equation}
%
Substituting \eqref{eq:RTComp}, \eqref{eq:RhoRHD}, \eqref{eq:Index}, and \eqref{eq:MagRMHD} into \eqref{eq:PressureODE} gives
\begin{align*}
	\exp(1-r^2)\left[2\Gamma\sigma\rho+(\Gamma-1)B_0^2(2- r^2)\right]= \left((\Gamma-1)\rho+\Gamma p\right)\dfrac{f^2}{1-r^2f^2},
\end{align*}
which implies
\begin{equation}\label{eq:f2}
	f^2=\dfrac{\kappa \exp(1-r^2)}{\kappa r^2\exp(1-r^2) + (\Gamma-1)\rho + \Gamma p},
\end{equation}
where
\begin{equation}\label{eq:kappa}
	\kappa := 2\Gamma \sigma\rho + (\Gamma-1)B_0^2(2-r^2).
\end{equation}
It is obvious that both $\rho=(1-\sigma \exp(1-r^2))^{\frac{1}{\Gamma-1}}$ and $p=\rho^\Gamma$ are  monotone increasing in $r\in[0,+\infty)$, so that the minimum of $(\Gamma-1)\rho + \Gamma p$ is $(\Gamma-1)(1-\sigma \exp(1))^{\frac{1}{\Gamma-1}} + \Gamma(1-\sigma \exp(1))^\Gamma$, which is strictly positive.
Moreover, if $\kappa\geqslant0$, then $f$ is well defined.
Through some search, it is found that with the adiabatic index $\Gamma=5/3$, when
\begin{equation}\label{eq:C1B0}
	\sigma=0.2,~B_0=0.05,
\end{equation}
$\kappa$ is positive for $r<20$, which is enough for the setup of our test problem
(the maximum $r$ in the 2D and 3D transformed domains in the rest frame, see the diamond domains in Figures \ref{fig:domain} and \ref{fig:3Ddomain}, are $10$ and $25\sqrt{2}/3$, respectively).
%
%\begin{figure}[!ht]
%\centering
%\includegraphics[width=0.6\textwidth]{./images/f2.pdf}
%\caption{Profiles of $\kappa$ and $f^2$ against $r$ with $\sigma=0.2$ and $B_0=0.1$.}
%\label{fig:f2}
%\end{figure}
%
%It can also be found that $f < 1.16\times 10^{-16}$ and $\abs{\bm{B}} < 1.51\times 10^{-8}$ when $r>6$, respectively, hence they will not cause boundary effects.
It can also be found that the velocities and magnetic fields vanish rapidly, hence they will not cause boundary effects.
Finally, the steady solution is determined by \eqref{eq:RhoPre}, \eqref{eq:VelMag}, \eqref{eq:RhoRHD}-\eqref{eq:Index}, \eqref{eq:MagRMHD}-\eqref{eq:C1B0}.

\begin{figure}[!ht]
	\centering
	\begin{subfigure}[b]{0.24\textwidth}
		\centering
		\includegraphics[width=1.0\textwidth]{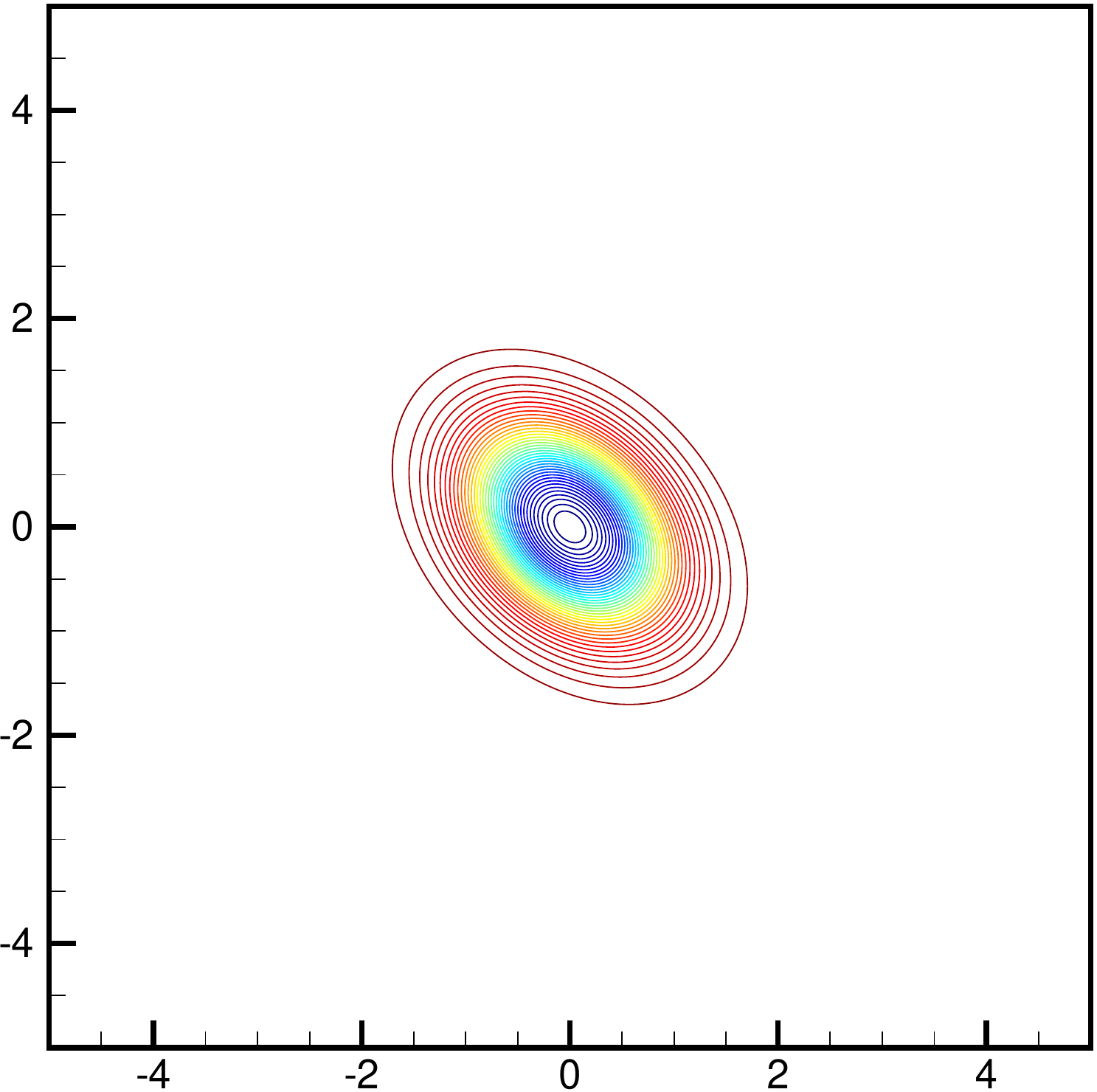}
		\caption{$\rho$}
	\end{subfigure}
	\begin{subfigure}[b]{0.24\textwidth}
		\centering
		\includegraphics[width=1.0\textwidth]{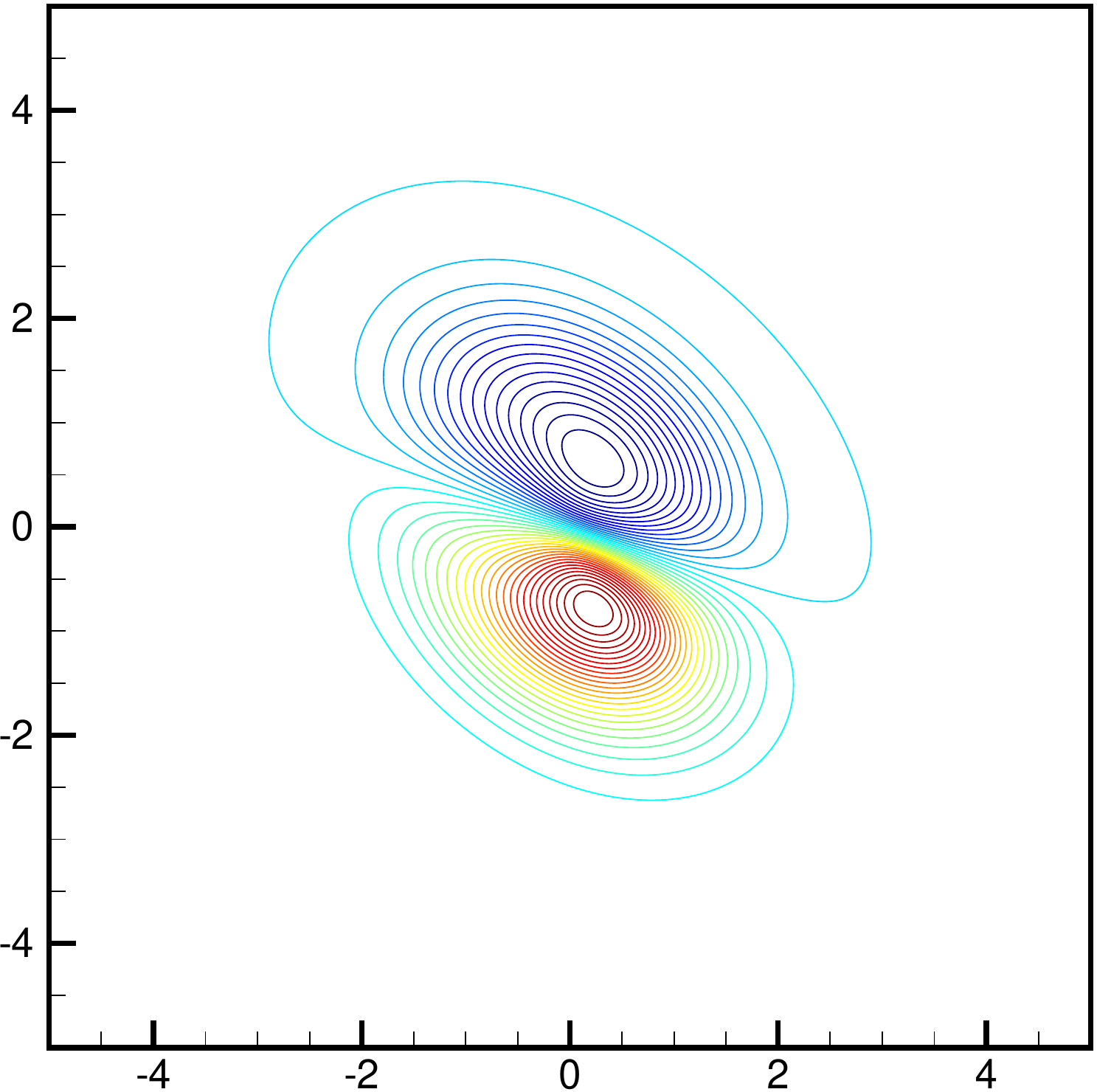}
		\caption{$\vx$}
	\end{subfigure}
	\begin{subfigure}[b]{0.24\textwidth}
		\centering
		\includegraphics[width=1.0\textwidth]{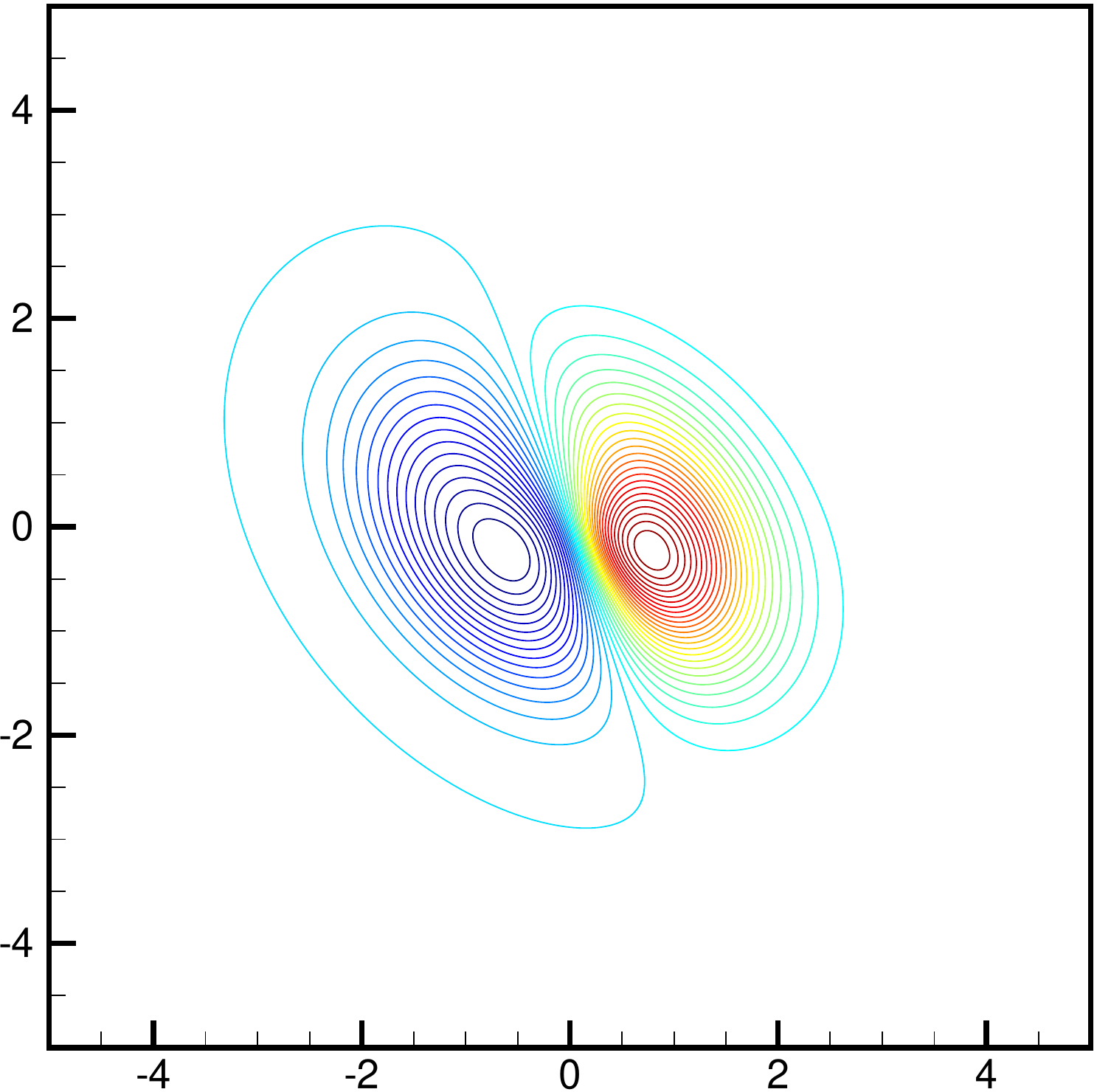}
		\caption{$\vy$}
	\end{subfigure}
	\begin{subfigure}[b]{0.24\textwidth}
		\centering
		\includegraphics[width=1.0\textwidth]{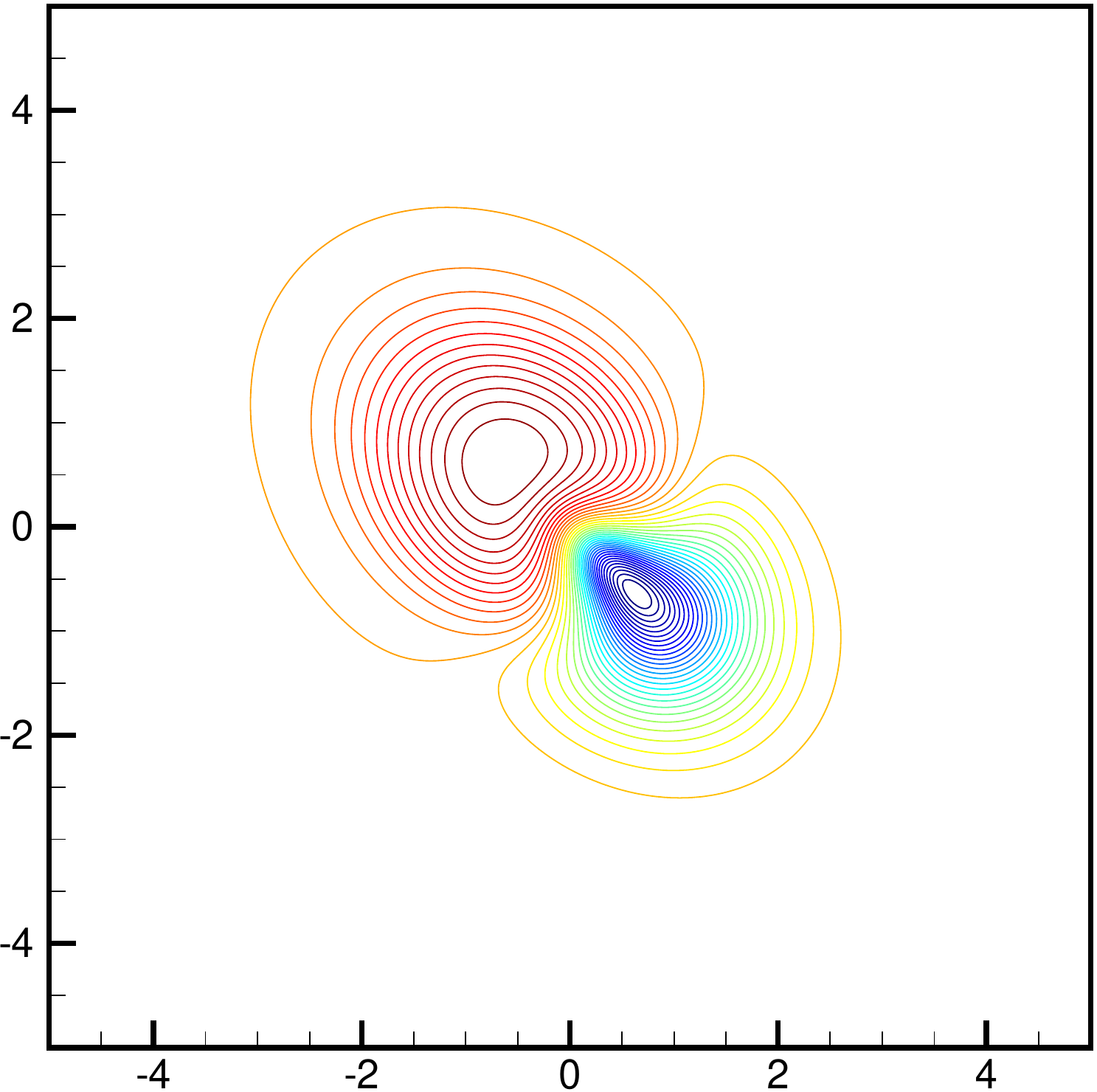}
		\caption{$\abs{\bv}$}
	\end{subfigure}

	\begin{subfigure}[b]{0.24\textwidth}
		\centering
		\includegraphics[width=1.0\textwidth]{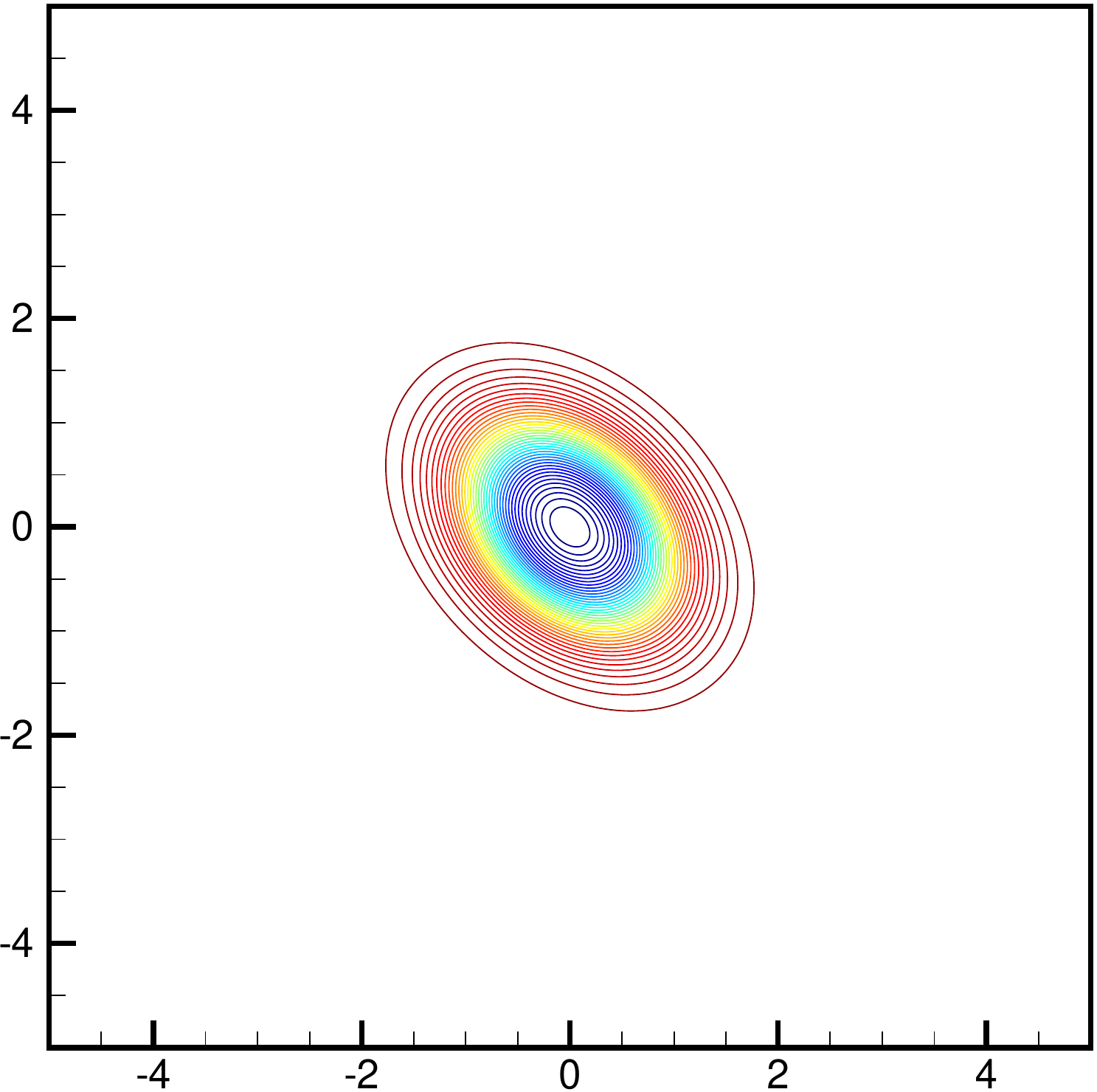}
		\caption{$p$}
	\end{subfigure}
	\begin{subfigure}[b]{0.24\textwidth}
		\centering
		\includegraphics[width=1.0\textwidth]{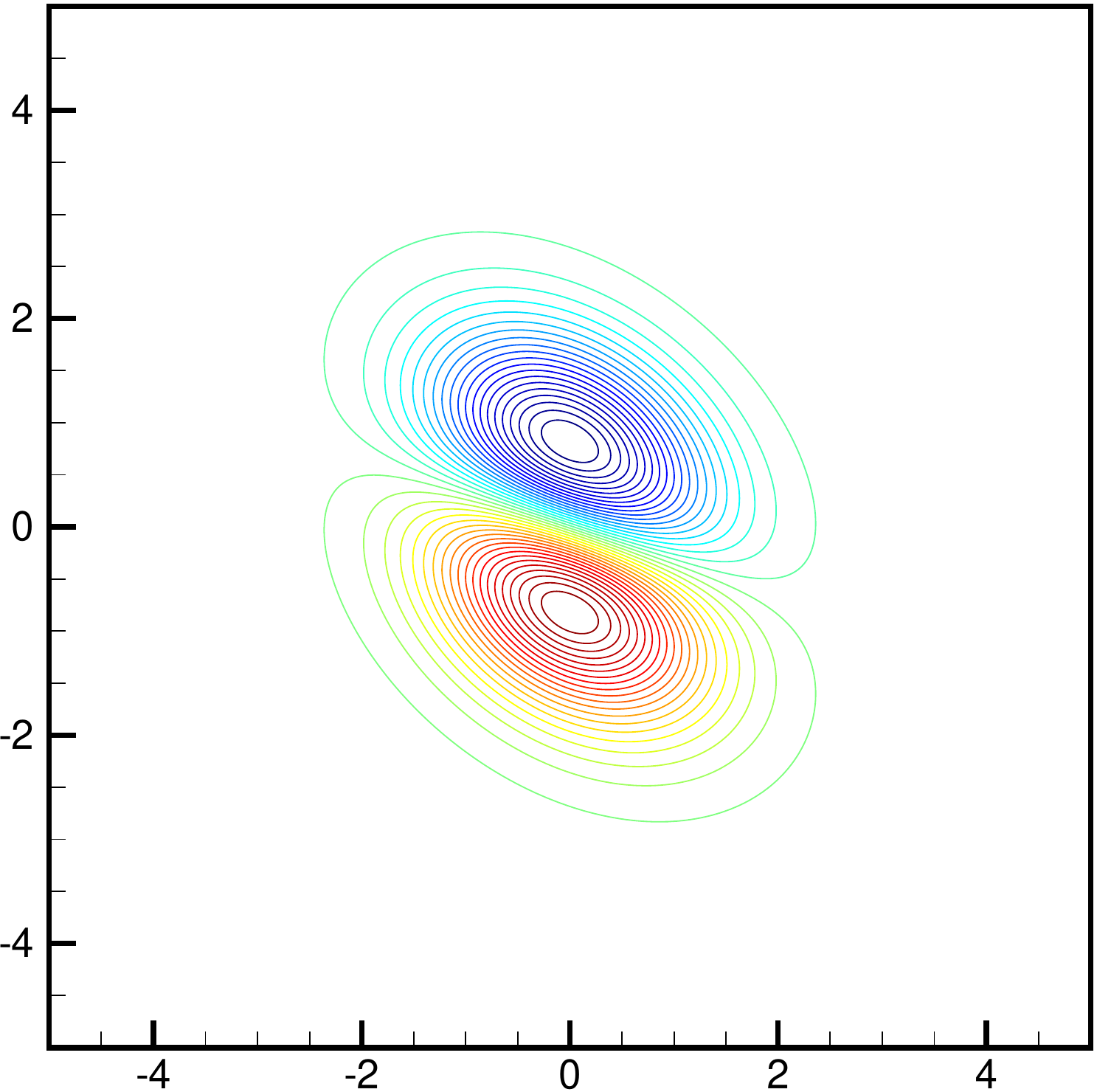}
		\caption{$\Bx$}
	\end{subfigure}
	\begin{subfigure}[b]{0.24\textwidth}
		\centering
		\includegraphics[width=1.0\textwidth]{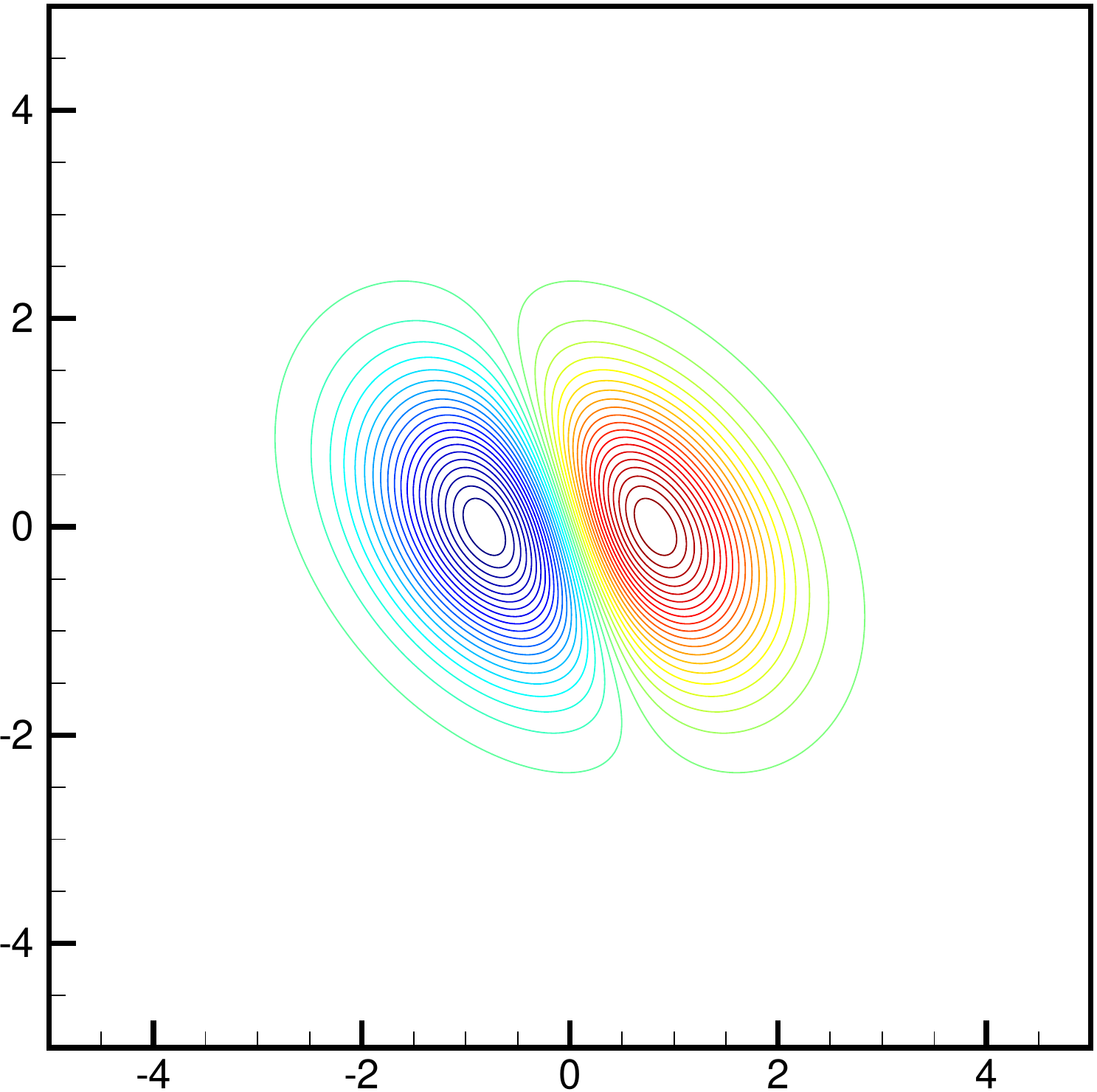}
		\caption{$\By$}
	\end{subfigure}
	\begin{subfigure}[b]{0.24\textwidth}
		\centering
		\includegraphics[width=1.0\textwidth]{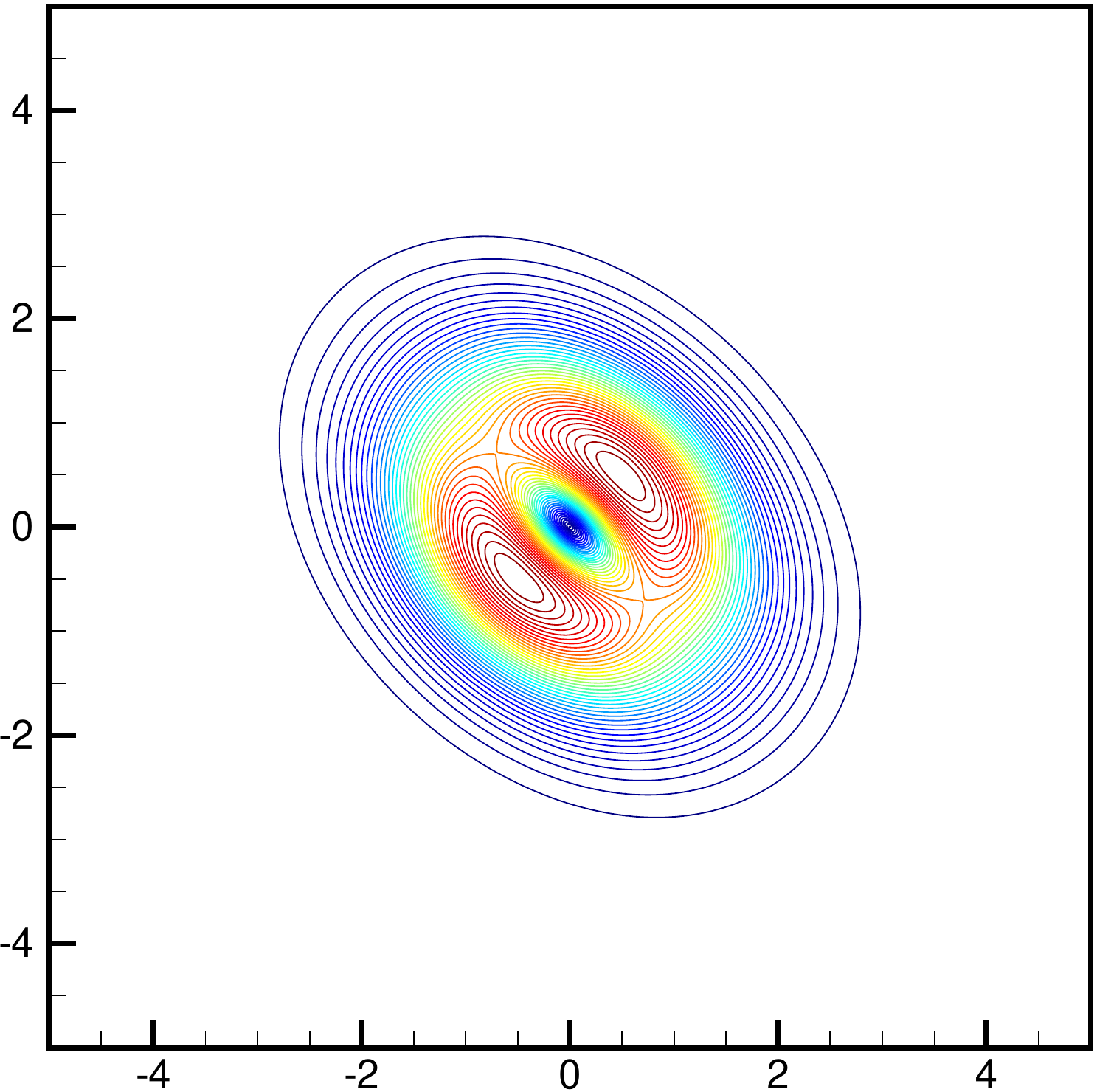}
		\caption{$\abs{\bm{B}}$}
	\end{subfigure}
	\caption{2D RMHD case: $40$ equally spaced contour lines of the initial solutions.}
	\label{fig:2DVortex}
\end{figure}

For the convenience of the readers, the specific expressions of the analytical solutions with $\bm{w}=(-0.5,-0.5)$ is listed here.
The analytical solution at time $t$ and the spatial point $(x_1,x_2)$ in the computational domain $[-R,R]\times[-R,R]$ with $R=5$ and the periodic boundary conditions  are given by
\begin{equation}\label{eq:Vortex1}
\begin{aligned}
	\rho&=(1-\sigma\exp(1-r^2))^{\frac{1}{\Gamma-1}},~ p=\rho^\Gamma,\\
	\bm{v}&=\frac{1}{4-2(\widetilde{v}_1+\widetilde{v}_2)}((2+\sqrt{2})\widetilde{v}_1+(2-\sqrt{2})\widetilde{v}_2-2,
	~(2+\sqrt{2})\widetilde{v}_2+(2-\sqrt{2})\widetilde{v}_1-2,~0),\\
	\bm{B}&=\frac12\left((\sqrt{2}+1)\widetilde{B}_1 - (\sqrt{2}-1)\widetilde{B}_2,
	~(\sqrt{2}+1)\widetilde{B}_2 - (\sqrt{2}-1)\widetilde{B}_1,~0\right),
\end{aligned}
\end{equation}
where
\begin{equation}\label{eq:Vortex2}
\begin{aligned}
	&\Gamma=5/3,~\sigma=0.2,~B_0=0.05,~r=\sqrt{\widetilde{x}_1^2+\widetilde{x}_2^2},\\
	&\widetilde{x}_k=\widehat{x}_k+(\sqrt{2}-1)(\widehat{x}_1 + \widehat{x}_2)/2,~k=1,2,\\
	&(\widehat{x}_1,\widehat{x}_2)=(2k_1R + x_1 + t/2,~2k_2R + x_2 + t/2), ~(\widehat{x}_1,\widehat{x}_2)\in [-R,R]\times[-R,R], ~k_1,k_2\in\mathbb{Z},\\
	& (\widetilde{v}_1,\widetilde{v}_2)=(-\widetilde{x}_2,\widetilde{x}_1)f,~ f=\sqrt{\dfrac{\kappa \exp(1-r^2)}{\kappa r^2\exp(1-r^2) + (\Gamma-1)\rho + \Gamma p}},~
	\kappa = 2\Gamma \sigma\rho + (\Gamma-1)B_0^2(2-r^2),	\\
	& (\widetilde{B}_1,\widetilde{B}_2)=B_0\exp(1-r^2)(-\widetilde{x}_2,\widetilde{x}_1).
\end{aligned}
\end{equation}
Setting $\bm{B}=\bm{0}$ gives the solution of the isentropic vortex problem for the RHD case.
In order to understand them intuitively,
Figure \ref{fig:2DVortex} displays the initial solutions \eqref{eq:Vortex1}-\eqref{eq:Vortex2} in the computational domain.
The initial rest-mass density $\rho$ is also plotted in both the rest frame and moving frame, see Figure \ref{fig:domain}, which clearly shows the Lorentz contraction.

\begin{rmk}\rm
The vortex problem in \cite{Ling2019Physical} corresponds to $\sigma=\dfrac{10.28(\Gamma-1)}{8\Gamma\pi^2}$, while $\sigma=\dfrac{25(\Gamma-1)}{8\Gamma\pi^2}$ in \cite{Duan2020RHD}.
\end{rmk}

\begin{rmk}\rm
	For the non-relativistic MHD case in \cite{Balsara2004Second}, the vortex problem is obtained by setting $\rho=1$, and then solving the pressure $p$ from \eqref{eq:PressureODE}, so that it is not an isentropic vortex.
\end{rmk}

Before ending this section, the  isentropic vortex problems with the solutions \eqref{eq:Vortex1}-\eqref{eq:Vortex2} are solved by the existing numerical schemes to verify the correctness.
The spatial discretizations  are  the $6$th-order accurate entropy conservative finite difference schemes for  the RHD equations \cite{Duan2020RHD} and  the RMHD equations \cite{Wu2020Entropy}, respectively, while the third-order accurate strong-stability-preserving Runge-Kutta (SSP-RK3) scheme is used in time.
A series of $N\times N$ meshes with the spatial mesh stepsize $h=2R/N$ are used, and the time stepsize is chosen as $\Delta t=h^2$ to make the spatial errors dominant.
The output time is $t=20$, so that the vortex travels and returns to the original position after a period.
Table \ref{tab:2DVortex} lists the errors defined by
\begin{equation}\label{eq:error2d}
	\varepsilon_{1} = \sum_{i,j=1}^{N}\sum_{s} \Delta \bm{W}_s/N^2,~
	\varepsilon_{2} = \sqrt{\sum_{i,j=1}^{N}\sum_{s} \Delta \bm{W}_s^2/N^2},~
	\varepsilon_{\infty} = \max_{i,j=1}^{N}\max_{s} \Delta \bm{W}_s,
\end{equation}
where $\Delta \bm{W}_s=\abs{(\bW_h)_s-\bW_s}$, and $\bW_h$ denotes the numerical approximation to the the exact solutions $\bW=(\rho,\vx,\vy,p,\Bx,\By)^\mathrm{T}$.
It can be seen that $6$th-order accuracy is obtained.

\begin{table}[!ht]
	\centering
	{\footnotesize
		\begin{tabular}{r|cc|cc|cc|cc|cc|cc}
			\toprule
			& \multicolumn{6}{c|}{RHD} & \multicolumn{6}{c}{RMHD} \\ \midrule
			$N$ & $\varepsilon_{1}$ & order & $\varepsilon_{2}$ & order & $\varepsilon_{\infty}$ & order & $\varepsilon_{1}$ & order & $\varepsilon_{2}$ & order & $\varepsilon_{\infty}$ & order  \\ \midrule
			40 & 2.57e-03 &  -   & 3.36e-03 &  -   & 1.95e-02 &  -   & 3.53e-03 &  -   & 3.85e-03 &  -   & 1.90e-02 &  -   \\
			80 & 2.97e-05 & 6.43 & 4.87e-05 & 6.11 & 3.60e-04 & 5.76 & 5.07e-05 & 6.12 & 7.27e-05 & 5.73 & 6.42e-04 & 4.89 \\
			120 & 2.57e-06 & 6.04 & 4.41e-06 & 5.92 & 3.37e-05 & 5.84 & 4.40e-06 & 6.03 & 7.01e-06 & 5.77 & 6.94e-05 & 5.49 \\
			160 & 4.62e-07 & 5.97 & 7.95e-07 & 5.96 & 5.58e-06 & 6.25 & 7.81e-07 & 6.01 & 1.30e-06 & 5.85 & 1.46e-05 & 5.43 \\
			200 & 1.26e-07 & 5.81 & 2.11e-07 & 5.96 & 1.44e-06 & 6.08 & 2.11e-07 & 5.87 & 3.49e-07 & 5.91 & 3.75e-06 & 6.08 \\
			\bottomrule
		\end{tabular}
	}
	\caption{2D isentropic vortex problem: Errors and orders of convergence at $t=20$.}
	\label{tab:2DVortex}
\end{table}

\begin{figure}[!ht]
	\centering
	\begin{subfigure}[b]{0.48\textwidth}
		\centering
		\includegraphics[width=1.0\textwidth]{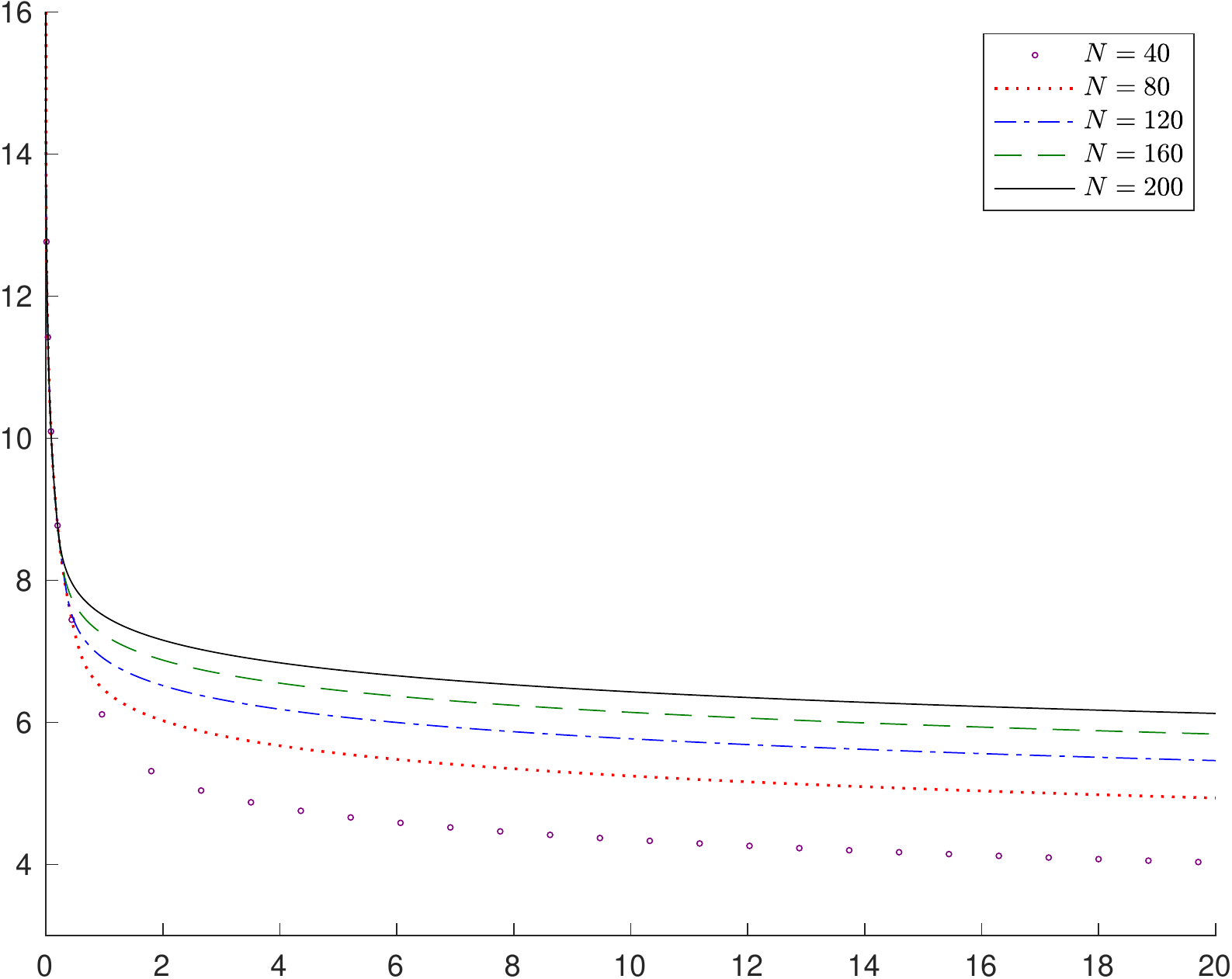}
		\caption{$\Delta t = h$}
	\end{subfigure}
	\begin{subfigure}[b]{0.48\textwidth}
		\centering
		\includegraphics[width=1.0\textwidth]{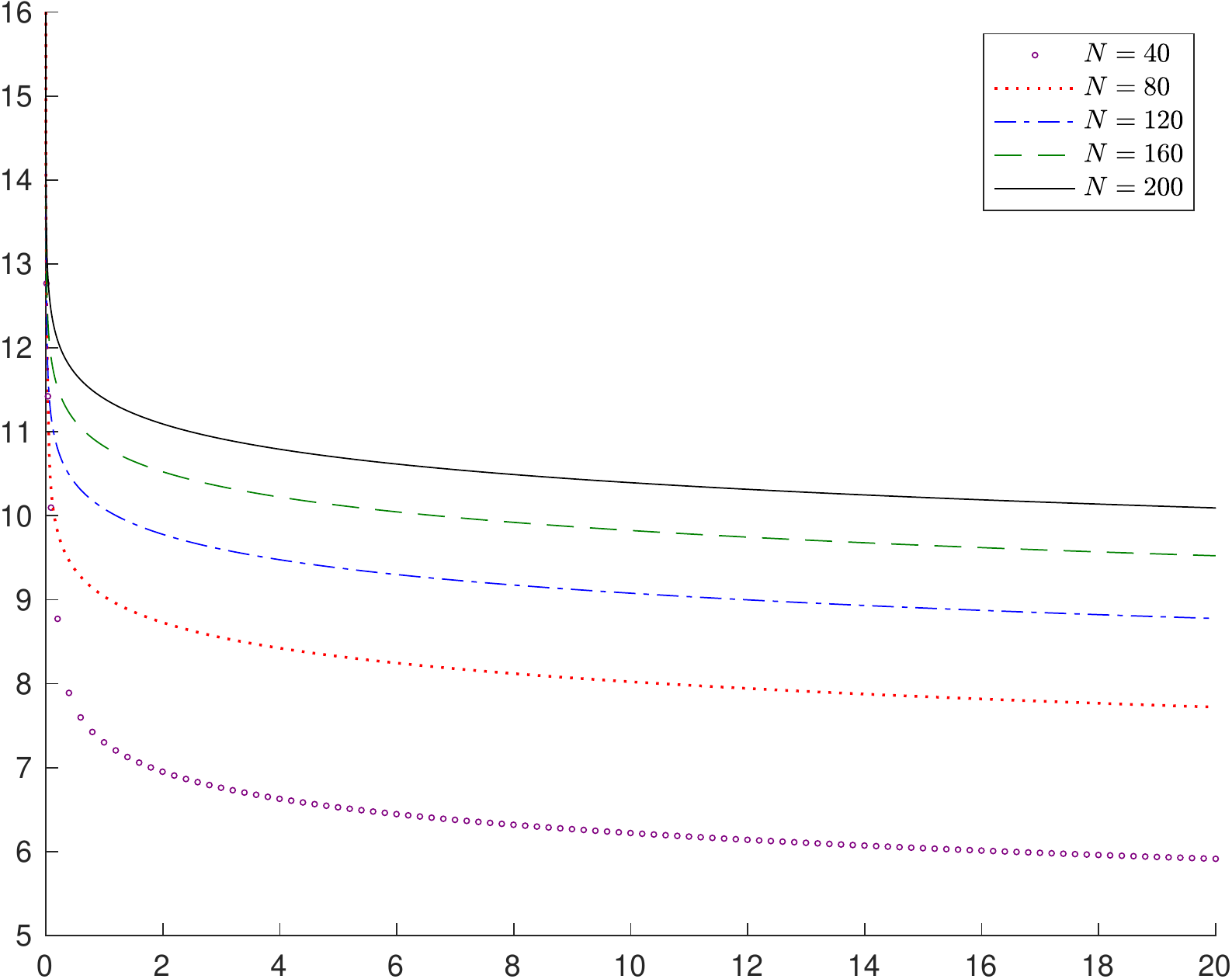}
		\caption{$\Delta t = h^2$}
	\end{subfigure}
	\caption{2D RMHD case: The evolution of $-\log_{10}(-\eta_h)$ on different meshes.}
	\label{fig:TotalEntropy}
\end{figure}

To verify the isentropic property of the problem, define the discrete total entropy by
\begin{equation*}
	\eta_h := \sum_{i,j}^N \eta((\bW_h)_{ij}) /N^2,
\end{equation*}
where $\eta=-\rho W\left(\ln p - \Gamma\ln \rho\right)$.
Figure \ref{fig:TotalEntropy} presents the evolution of $\eta_h$ for the RMHD case, and it can be seen that $\eta_h$ decays as time increases, and converges as $h$ decreases.
To further check the convergence order of $\eta_h$, Table \ref{tab:2DTotalEntropy}  gives $\eta_h$ on different meshes obtained with $\Delta t=h$ and $\Delta t=h^2$, which indicates that $\eta_h$ decays with $\mathcal{O}\left(\Delta t^3\right)$ when the SSP-RK3 is used, similar to the results in \cite{Duan2020RMHD}.

\begin{table}[!ht]
	\centering
	\begin{tabular}{r|cc|cc|cc|cc}
		\toprule
		& \multicolumn{4}{c|}{RHD} & \multicolumn{4}{c}{RMHD} \\ \midrule
		& \multicolumn{2}{c|}{$\Delta t=h$} & \multicolumn{2}{c|}{$\Delta t=h^2$} & \multicolumn{2}{c|}{$\Delta t=h$} & \multicolumn{2}{c}{$\Delta t=h^2$}\\ \midrule
		$h$ & $\eta_h$ & order & $\eta_h$ & order & $\eta_h$ & order & $\eta_h$ & order \\ \midrule
		$1/4$ &   -8.93e-05 &  -   & -1.16e-06 &  -       & -9.39e-05 &  -              & -1.22e-06 &  -     \\
		$1/8$ &  -1.10e-05 & 3.02 & -1.80e-08 & 6.01   & -1.16e-05 & 3.02           & -1.91e-08 & 6.00 \\
		$1/12$ &-3.27e-06 & 2.99 & -1.58e-09 & 6.00   & -3.45e-06 & 2.99          & -1.68e-09 & 6.00 \\
		$1/16$ &-1.38e-06 & 2.99 & -2.82e-10 & 6.00   & -1.46e-06 & 2.99          & -3.00e-10 & 5.98 \\
		$1/20$ & -7.10e-07 & 3.00 & -7.45e-11 & 5.97    & -7.49e-07 & 2.99         & -8.07e-11 & 5.88 \\
		\bottomrule
	\end{tabular}
	\caption{2D isentropic vortex problem: The discrete total entropies $\eta_h$ and  corresponding convergence orders at $t=20$.}
	\label{tab:2DTotalEntropy}
\end{table}

\section{Extension to the 3D case}
\begin{figure}[!ht]
	\centering
	\includegraphics[width=0.42\textwidth]{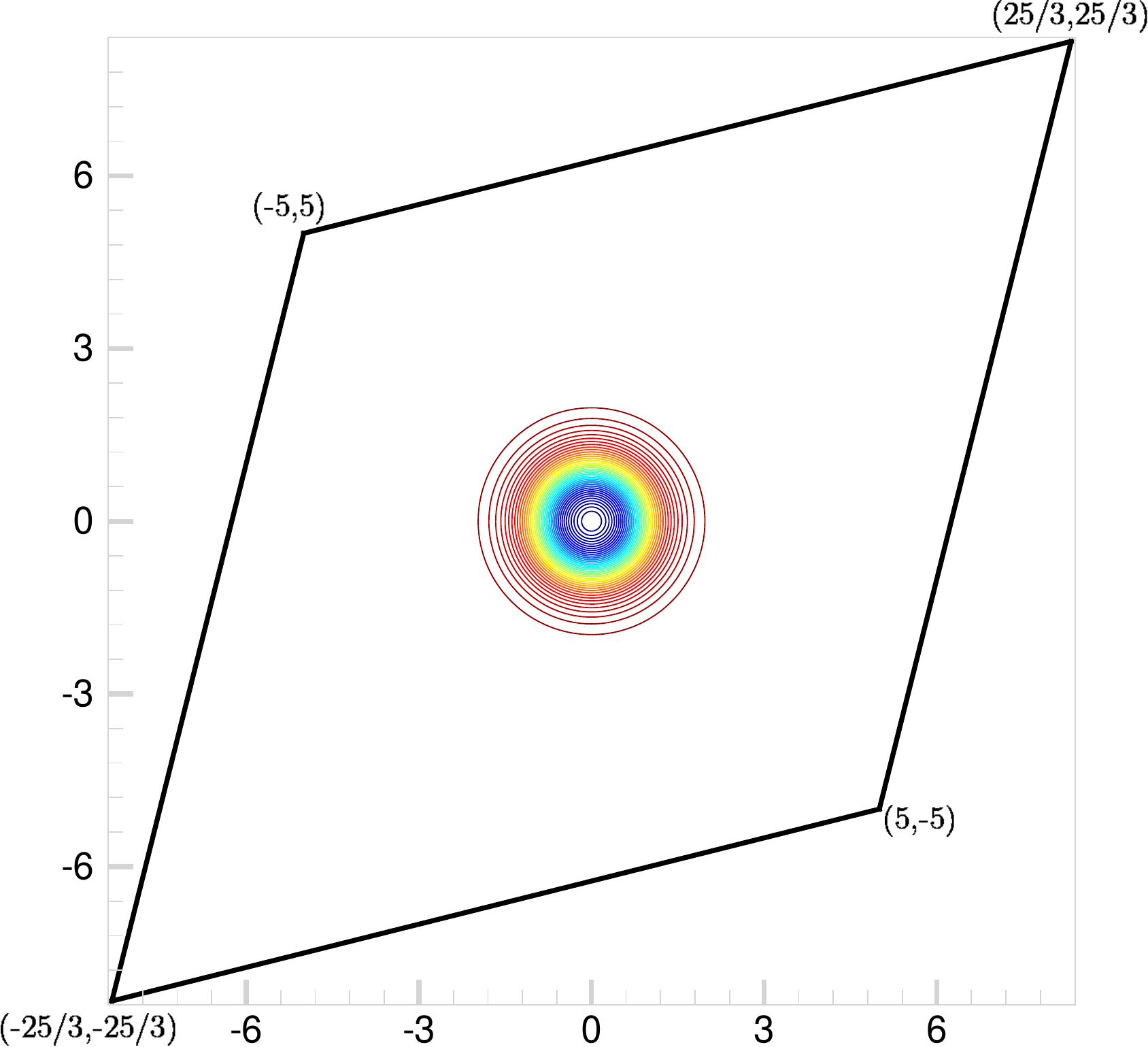}
	\caption{3D RMHD case: $40$ equally spaced contour lines of the initial $\rho$ in $\Omega_0$ in $\widetilde{x}_1-\widetilde{x}_2$ plane in the rest frame with $R=5$ and $\bm{w}=(-0.5,-0.5,-0.5)$.}
	\label{fig:3Ddomain}
\end{figure}

The 2D vortex presented in the last section can be viewed as a slice of a 3D cylindrical vortex, in which all the slices in the $x_3$-direction are the same.
This section considers the 3D case by rotating the cylindrical vortex to the diagonal of a cuboid computational domain $[-R,R]\times[-R,R]\times[-aR,aR]$, with $a$ to be determined later, so that the primitive variables are not constant in the $x_3$-direction.
It is achieved by using the following Lorentz transformation
\begin{equation*}
	\begin{pmatrix}
		\widetilde{t} \\ \widetilde{x}_1 \\ \widetilde{x}_2 \\ \widetilde{x}_3 \\
	\end{pmatrix}
	=\begin{pmatrix}
		\gamma & -\gamma w_1 & -\gamma w_2 & -\gamma w_3 \\
		-\gamma w_1 & 1+(\gamma-1) w_1^2/\abs{\bm{w}}^2 & (\gamma-1) w_1w_2/\abs{\bm{w}}^2 & (\gamma-1) w_1w_3/\abs{\bm{w}}^2 \\
		-\gamma w_2 & (\gamma-1) w_1w_2/\abs{\bm{w}}^2 & 1+(\gamma-1) w_2^2/\abs{\bm{w}}^2 & (\gamma-1) w_2w_3/\abs{\bm{w}}^2 \\
		-\gamma w_3 & (\gamma-1) w_1w_3/\abs{\bm{w}}^2 & (\gamma-1) w_2w_3/\abs{\bm{w}}^2 & 1+(\gamma-1) w_3^2/\abs{\bm{w}}^2 \\
	\end{pmatrix}
	\begin{pmatrix}
		t \\ x_1 \\ x_2 \\ x_3 \\
	\end{pmatrix},
\end{equation*}
where $\bm{w}=(w_1,w_2,w_3)$ is the velocity vector of the coordinate system $\widetilde{S}$ relative to the coordinate system $S$.
Choose $w_1=w_2=w_3=-0.5$ such that the Lorentz factor $\gamma=1/\sqrt{1-\abs{\bm{w}}^2}=2$.
To make the periodic boundary conditions work, $(t,x_1,x_2,x_3)=(0,-R,-R,aR)$ should be transformed to $(\widetilde{t}^*,0,0,\widetilde{x}_3^*)$ in the rest frame with arbitrary $\widetilde{t}^*$ and $\widetilde{x}_3^*$.
It is equivalent to $-4R/3-R/3+aR/3=0$, i.e. $a=5$.
After transformed into the rest frame $(\widetilde{x}_1,\widetilde{x}_2,\widetilde{x}_3)$, the solution is periodic in the $\widetilde{x}_1-\widetilde{x}_2$ plane, and the period region is a diamond  $\Omega_0$,
% with four vertices $(25/3,25/3),(5,-5),(-25/3,-25/3),(-5,5)$
 see Figure \ref{fig:3Ddomain}.

\begin{figure}[!ht]
	\centering
	\begin{subfigure}[b]{0.48\textwidth}
		\centering  %\includegraphics[width=15cm,height=6cm]
		\includegraphics[width=7cm,height=6cm, trim=90 0 90 0, clip]
      {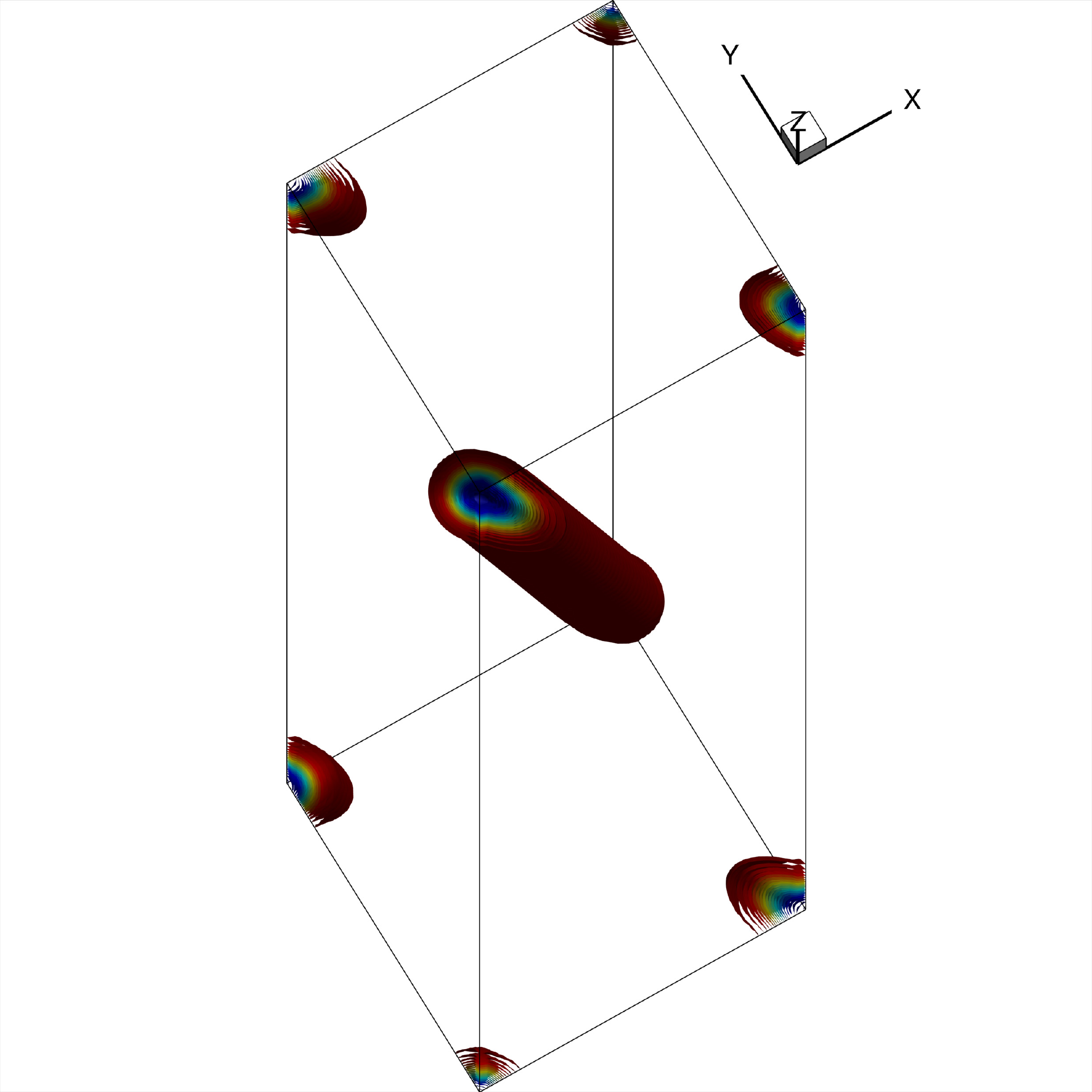}
		\caption{$\rho$}
	\end{subfigure}
	\begin{subfigure}[b]{0.48\textwidth}
		\centering
		\includegraphics[width=7cm,height=6cm, trim=90 0 90 0, clip]
    {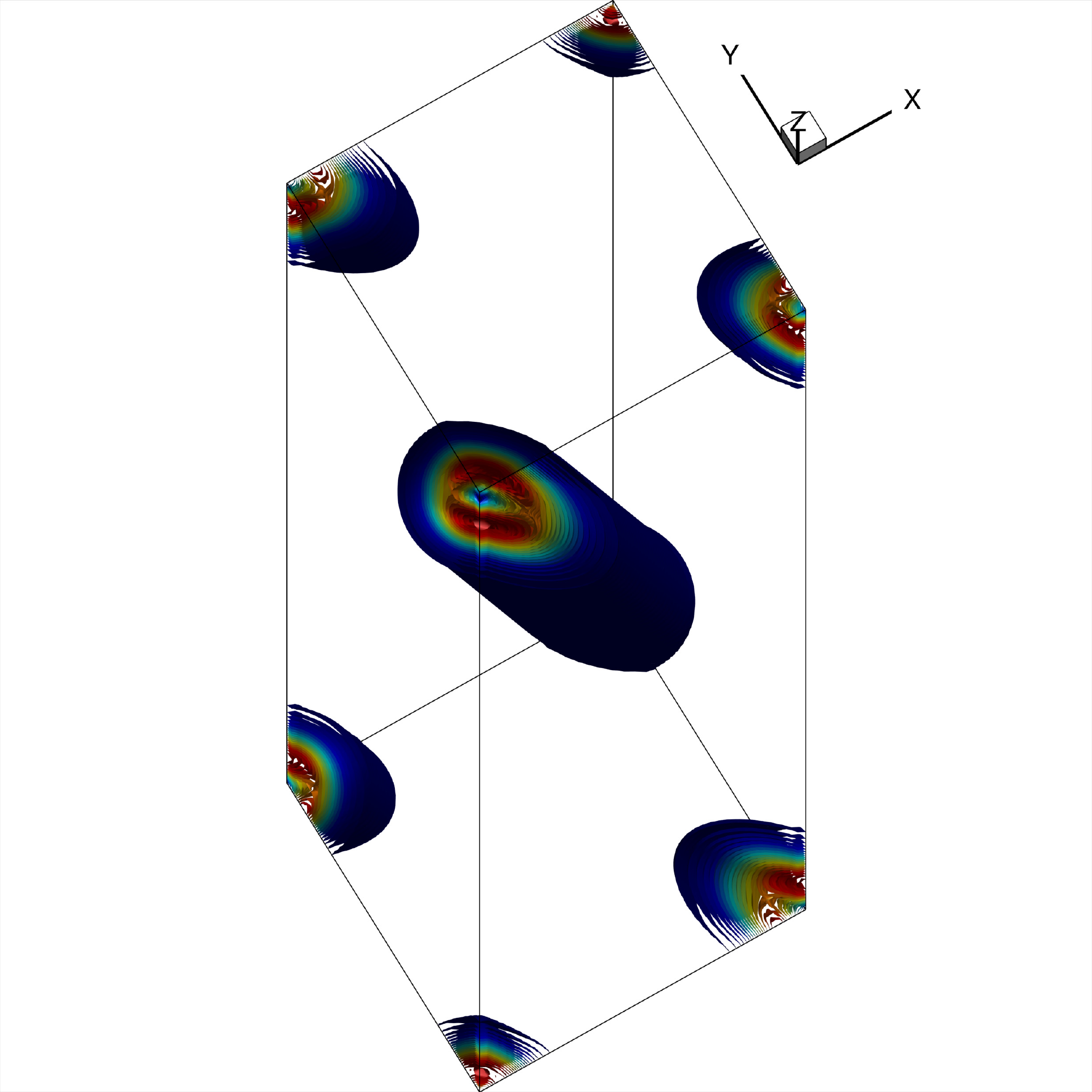}
		\caption{$\abs{\bm{B}}$}
	\end{subfigure}
	\caption{3D RMHD case: The $20$ equally spaced iso-surfaces of $\rho $ and $\abs{\bm{B}}$ at $t=0$.}
	\label{fig:3DVortex}
\end{figure}

For the convenience of the readers, the specific expressions of the analytical solutions are listed here.
The analytical solutions at  time $t$ and the spatial point $(x_1,x_2,x_3)$ in the computational domain $[-R,R]\times[-R,R]\times[-5R,5R]$ with $R=5$ and the periodic boundary conditions can be given by
\begin{equation}\label{eq:3DVortex1}
	\begin{aligned}
		\rho&=(1-\sigma\exp(1-r^2))^{\frac{1}{\Gamma-1}},~ p=\rho^\Gamma,\\
		\bm{v}&=\frac{1}{6-3(\widetilde{v}_1+\widetilde{v}_2)}(4\widetilde{v}_1+\widetilde{v}_2-3,
		~4\widetilde{v}_2+\widetilde{v}_1-3, ~\widetilde{v}_1+\widetilde{v}_2-3),\\
		\bm{B}&=\frac13\left(5\widetilde{B}_1 - \widetilde{B}_2,
		~5\widetilde{B}_2 - \widetilde{B}_1,
		~-\widetilde{B}_1 - \widetilde{B}_2\right),
	\end{aligned}
\end{equation}
where
\begin{equation}\label{eq:3DVortex2}
	\begin{aligned}
		&\Gamma=5/3,~\sigma=0.2,~B_0=0.05,~r=\sqrt{\widetilde{x}_1^2+\widetilde{x}_2^2},\\
		&(\widetilde{x}_1,\widetilde{x}_2)=(40/3k_1 + 10/3k_2 + \widehat{x}_1,~10/3k_1 + 40/3k_2 + \widehat{x}_2), ~(\widetilde{x}_1,\widetilde{x}_2)\in \Omega_0, ~k_1,k_2\in\mathbb{Z},\\
		&\widehat{x}_k=x_k+({x}_1 + {x}_2 + {x}_3)/3+t,~k=1,2,3,\\
		&(\widetilde{v}_1,\widetilde{v}_2)=(-\widetilde{x}_2,\widetilde{x}_1)f,~ f=\sqrt{\dfrac{\kappa \exp(1-r^2)}{\kappa r^2\exp(1-r^2) + (\Gamma-1)\rho + \Gamma p}},~
		\kappa = 2\Gamma \sigma\rho + (\Gamma-1)B_0^2(2-r^2),	\\
		& (\widetilde{B}_1,\widetilde{B}_2)=B_0\exp(1-r^2)(-\widetilde{x}_2,\widetilde{x}_1).
	\end{aligned}
\end{equation}
Figure \ref{fig:3DVortex} shows the $20$ equally spaced iso-surfaces of  $\rho $ and $\abs{\bm{B}}$ at $t=0$.
If setting $\bm{B}=\bm{0}$, then the solution of the 3D isentropic vortex problem for the RHD case is obtained.

\begin{rmk}\rm
	To compute $\widetilde{x}_1,\widetilde{x}_2$, one can proceed as follows.
	Let $(\widehat{x}_1,\widehat{x}_2)=(25/3a_1+5/3a_2, 5/3a_1+20/3a_2)$, then
	$a_1=(4\widehat{x}_1 - \widehat{x}_2)/25,~a_2=(4\widehat{x}_2 - \widehat{x}_1)/25$.
	Find $b_1,b_2\in[-1,1]$, such that $b_1=a_1 + 2m_1, ~b_2=a_2 + 2m_2, ~m_1,m_2\in\mathbb{Z}$,
	so that $(\widetilde{x}_1,\widetilde{x}_2)=(20/3b_1 + 5/3b_2, 5/3b_1 + 20/3b_2)$.
\end{rmk}

The 3D RHD and RMHD vortex problems are also computed by the 3D entropy conservative finite difference schemes being similar to the 2D case. The errors $\varepsilon_1$, $\varepsilon_2$, and $\varepsilon_\infty$, defined similarly in \eqref{eq:error2d}, and the convergence orders are listed in Table \ref{tab:3DVortex}.
It can be seen that the $6$th-order convergence is obtained.

\begin{table}[!ht]
	\centering
	{\footnotesize
		\begin{tabular}{r|cc|cc|cc|cc|cc|cc}
			\toprule
			& \multicolumn{6}{c|}{RHD} & \multicolumn{6}{c}{RMHD} \\ \midrule
			$N$ & $\varepsilon_{1}$ & order & $\varepsilon_{2}$ & order & $\varepsilon_{\infty}$ & order & $\varepsilon_{1}$ & order & $\varepsilon_{2}$ & order & $\varepsilon_{\infty}$ & order  \\ \midrule
			40 & 3.51e-05 &  -   & 8.39e-05 &  -   & 8.56e-04 &  -   & 3.74e-05 &  -   & 8.40e-05 &  -   & 8.49e-04 &  -   \\
			80 & 7.81e-07 & 5.49 & 1.98e-06 & 5.41 & 2.41e-05 & 5.15 & 8.57e-07 & 5.45 & 2.01e-06 & 5.38 & 2.42e-05 & 5.13 \\
			120 & 7.42e-08 & 5.81 & 1.90e-07 & 5.77 & 2.43e-06 & 5.67 & 8.25e-08 & 5.77 & 1.96e-07 & 5.74 & 2.44e-06 & 5.66 \\
			160 & 1.38e-08 & 5.84 & 3.51e-08 & 5.88 & 4.53e-07 & 5.83 & 1.55e-08 & 5.82 & 3.64e-08 & 5.85 & 4.67e-07 & 5.74 \\
			200 & 3.86e-09 & 5.71 & 9.40e-09 & 5.90 & 1.22e-07 & 5.88 & 4.33e-09 & 5.70 & 9.77e-09 & 5.89 & 1.30e-07 & 5.75 \\
			\bottomrule
		\end{tabular}
	}
	\caption{3D isentropic vortex problem: Errors and orders of convergence at $t=0.1$.}
	\label{tab:3DVortex}
\end{table}

\section{Conclusion}\label{section:Conclusion}
This note   provided the first analytical
solution of the 2D isentropic vortex problem with explicit algebraic expressions in the special relativistic hydrodynamics and magnetohydrodynamics and extended it to the 3D case.
It did not require any  ordinary differential equation solver, so that it would be useful and convenient for the code verification.

%% Acknowledgments %%%%%%%
\section*{Acknowledgments}

This work is financially supported by the National Key R\&D Program of China, Project Number 2020YFA0712000,
the Sino-German Cooperation Group Project (No. GZ 1465),
and High-performance Computing Platform of Peking University.

\end{document}